\documentclass[prd,aps,a4paper,superscriptaddress,twocolumn,nofootinbib,floatfix]{revtex4}
\pdfoutput=1
\usepackage{graphicx}
\usepackage{color}
\usepackage{dcolumn}
\usepackage{bm}
\usepackage{slashed}
\usepackage{amsmath}
\usepackage{latexsym}
\usepackage{amssymb}
\usepackage{mathrsfs}

\usepackage{placeins}

\usepackage{float}

\usepackage{threeparttable}
\usepackage{amsfonts}
\usepackage{wrapfig}
\usepackage{autobreak}

\allowdisplaybreaks
\usepackage[backref]{hyperref}
\hypersetup{colorlinks,			
	linkcolor=blue,
	citecolor=blue}
\allowdisplaybreaks
\begin{document}
\date{\today}
\title{Optimal TDI2.0  of sensitive curve for main space GW detector }
\author{Yu Tian,Zhi-Xiang Li}
\affiliation{School of Fundamental Physics and Mathematical Sciences, Hangzhou Institute for Advanced Study, UCAS, Hangzhou 310024, China}

\begin{abstract}
\quad  Time-delay interferometry (TDI) is a crucial technology for space-based gravitational wave detectors. Previous studies have identified the optimal TDI configuration for the first-generation. In this research, we used an Algebraic approach theory to describe the TDI space and employed a method to maximize the signal-to-noise ratio (SNR) to derive the optimal TDI combination for the second-generation. When this combination is used in the sensitivity curve, we observed enhancements of up to 1.91 times in the low-frequency domain and 2 to 3.5 times in the high-frequency domain compared to the Michelson combination. Furthermore, changes in the detector index significantly affect the optimization effect. We also present detection scenarios for several low-frequency gravitational wave sources. Compared to the first-generation TDI optimization, the SNR value for verification double white dwarfs (DWD) and the detection rate for DWD increase by 16.5$\%$.

\end{abstract}
\maketitle

\section{Introduction}
\quad The discovery of gravitational waves (GW) has greatly advanced the field of gravitational wave astronomy \cite{abbottObservationGravitationalWaves2016,PhysRevLett.116.061102}. It has led to the emergence of a new astronomy era that involves multi-band detection \cite{PhysRevD.93.021101, Accadia_2011, Harry_2010, Abbott_2009, Kawamura_2011}, multi-messenger detection, and various gravitational wave detectors. Furthermore, it has provided a new way to test theoretical models in various fields. Currently, gravitational wave detections cover the main frequency band above 1Hz, which includes pulsars and supernovae. Lower frequency gravitational waves in the range of $10^{-4}Hz-1Hz$ are detected using methods such as Extreme Mass Ratio Inspirals (EMRI) and compact binaries \cite{Belczynski_2010}. Meanwhile, supermassive black holes (SMBH) emit gravitational waves in the frequency band of $10^{-9}Hz-10^{-2}Hz$ \cite{Rodriguez_2006}. Gravitational wave detectors that operate in this frequency band include LISA \cite{Karsten.Danzmann_2003}, Taiji \cite{10.1093.nsr.nwx116}, and Tianqin \cite{Luo_2016}. Based on theoretical models of main GW sources, we can search for the optimal data processing methods for specific scientific objectives and detect different wave sources within the existing conditions to achieve maximum scientific satisfaction.

\quad Ground-based wave detectors such as LIGO and Virgo can suppress laser frequency noise to a very low level due to the minimal change of arm length. However, if the Michelson interference is placed in space to detect low-frequency sources, the ground gravity gradient noise level is too high to achieve the same effect \cite{PhysRevD.60.082001}. In space gravitational wave detectors, gravitational waves are detected through monitoring interference signals. The optical path between adjacent satellites constantly changes over time, leading to laser channel noise of an unacceptable magnitude \cite{PhysRevD.70.081101}. To suppress laser frequency noise, Time Delay Interferometry (TDI) is the main arithmetic \cite{Otto2015, AET99, amaro-seoaneLaserInterferometerSpace2017, babakLISASensitivitySNR2021, GeometricTDI}. Sensitivity curves are a useful tool to show the influence of different TDI configurations. The equal-arm configuration TDI1.0 \cite{TDI1.0} has been widely used in space gravitational wave detectors. However, the unequal-arm configuration TDI2.0 \cite{TDI2.0} is more suitable for real detection situations due to the movement of the satellites. This configuration can effectively reduce the impact of satellite drift and have a sufficiently small residual laser phase noise to extract gravitational waves. TDI X channel can suppress laser frequency noise by up to eight orders of magnitude. Compared to the X configuration, the existing first-generation optimal TDI channel combination, A, E, T \cite{TDI1.0opt, Algebraic.approach}, can be improved by $\sqrt {2}$ in low frequency and $\sqrt {3}$ in high frequency. Currently, K. Rajesh Nayak has generalized the corresponding TDI form for the known direction of the wave source \cite{TDI1.0opt.direction, optimal.LISA.sensitive}.

 \quad In this paper, we have obtained the optimal sensitivity by optimum weighting of second-generation Time Delay Interferometry (TDI) configurations, which results in an optimal signal-to-noise ratio (SNR) for detecting primary types of target gravitational wave sources. Compared to the TDI X channel, the second-generation optimal configuration can improve the sensitive curve by  up to 1.91 times under the low-frequency approximation and get 2 times even up to 3.5 times in the high-frequency approximation at certain frequency points . This improvement can effectively enhance the source detection rate and SNR. Moreover, using the code published by Ollie Burke and Andrea Antonelli, our proposed channel can improve the parameter estimation accuracy by up to 2 times in case of source confusion with deviation accumulation and by up to 10 times in the case of global fit.

\quad The paper is organized as follows. In Section II, we briefly introduce the basic process of the first-generation optimal TDI and explain the idea and development process behind the second-generation optimal TDI. In Section III, we obtain the PSD formula and sensitivity curve of the second-generation optimal TDI and compare it with other TDI configurations. In Section IV, we evaluate the SNR values and detection rates of the target wave source based on the optimal TDI configuration. In Section V, we apply parameter estimation.

\section{ TDI1.0 and TDI2.0  configuration of optimal SNR }

\begin{figure}[ht!]
\centering
\includegraphics[width=\columnwidth]{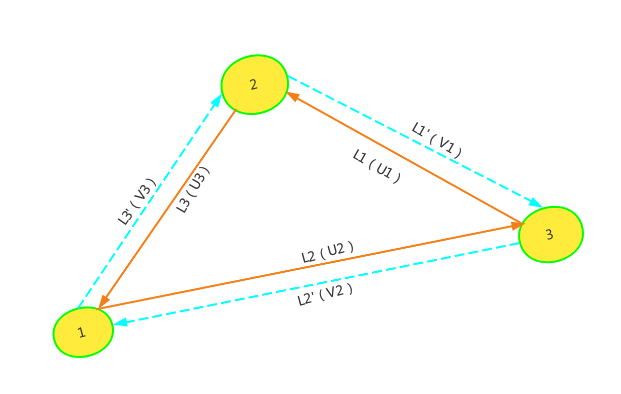}
\caption{Optical path configuration of space gravitational wave detector}
\label{fig:TDIFL}
\end{figure}

\quad In this section, we provide a brief overview of TDI, as necessary for the analysis presented in Fig. \ref{fig:TDIFL}. For space-based gravitational wave detectors, the Michelson interferometer is typically used on three satellites in orbit, labeled $SC_i$. The corresponding optical paths are denoted by L1, L2, and L3 ($L1'$, $L2'$, $L3'$) in counterclockwise (clockwise) order, and yield six basic data streams denoted by $U_i$ and $V_i$. The data is recorded by measuring the Doppler shift phase, including both noise and gravitational wave signals. In this work, we consider only the simplest case and ignore all other noises except shot noise and acceleration noise. The frequency fluctuation of the data stream is given by \cite{TDIdatastream}.

\begin{center}
    \begin{equation}
    \begin{aligned}
    C_{i}(t) = \frac{\triangle V_{i}(t)}{V_{0}(t)}
    \end{aligned}
    \label{eq:frequency fluctuation}
    \end{equation}
\end{center}

\quad where ${\triangle V_{i}(t)}$ is frequency fluctuation(${ V_{}(t)}$ is reference frequency) of the laser on $SC'_{i}$,t is gravitational wave travel time between adjacent SC.

\quad Laser frequency noise is a major source of interference that affects the accuracy of gravitational wave detection. To mitigate this effect, the main technology used is time delay interference (TDI), which involves time-delaying the data streams and recombining them linearly. This technique assumes that the data streams between the two optical benches (OBs) in each SC are equal, and introduces time delay interference data streams. The time delay interference operator $D_{j}C_{i}(t)$ is defined as $C_{i}(t-\frac{L_{j}}{c})$\cite{Tinto2014}, and the time delay interference data stream is expressed mathematically as follows:
\begin{equation}
\centering
\begin{split}
\label{time delay interference data stream}
U_{i}=D_{(i+1)}C_{(i+2)}-C_{i}\\
V_{i}=C_{i}-D_{(i+2)}C_{(i+1)}
\end{split}
\end{equation}

\quad Combine all the data streams, ideally eliminating all laser frequency noise, and get
\begin{equation}
\centering
\begin{split}
\label{sum of time delay interference data stream}
\sum_{i=1}^{3} p_{i} V^{i}+p_{i^{'}} U^{i}=0
\end{split}
\end{equation}

\quad where $p_{i}$ and $p_{i_{'}}$ is coefficient of time time delay interference operator.

\quad The current several major configurations of coefficient of $q_{i}$ and $q_{i^{'}}$ are  express in \cite{GeometricTDI}\cite{Algebraic.approach} \cite{Wangpanpan2021}.
So let's list the needed-configuration coefficients
\begin{equation}
\label{eq:TDIX coefficients}
\begin{aligned}
&\text{TDIX}_{1}\\&
P_{1}=\left(\mathcal{D}_{2^{\prime} 2}-1\right), P_{2}=0,P_{3}=\left(\mathcal{D}_{2^{\prime}}-\mathcal{D}_{33^{\prime} 2^{\prime}}\right) 
\\&P_{1^{\prime}}=\left(1-\mathcal{D}_{33^{\prime}}\right)
P_{2^{\prime}}=\left(\mathcal{D}_{2^{\prime} 23}-\mathcal{D}_{3}\right)
P_{3^{\prime}}=05\\
&\text{TDIX}_{2}\\&
P_{1}=-\left(1-\mathcal{D}_{2^{\prime} 2}-\mathcal{D}_{2^{\prime} 233^{\prime}}+\mathcal{D}_{33^{\prime} 2^{\prime} 2^{\prime} 2}\right),P_{2}=0 \\&
P_{3}=\left(1-\mathcal{D}_{33^{\prime}}-\mathcal{D}_{33^{\prime} 2^{\prime} 2}+\mathcal{D}_{2^{\prime} 233^{\prime} 33^{\prime}}\right) \mathcal{D}_{2^{\prime}}\\&
P_{1^{\prime}}=\left(1-\mathcal{D}_{33^{\prime}}-\mathcal{D}_{33^{\prime} 2^{\prime} 2}+\mathcal{D}_{2^{\prime} 233^{\prime} 33^{\prime}}\right) \\&
P_{2 \prime}=-\left(1-\mathcal{D}_{2^{\prime} 2}-\mathcal{D}_{2^{\prime} 233^{\prime}}+\mathcal{D}_{33^{\prime} 2^{\prime} 22^{\prime} 2}\right), P_{3 \prime}=0 \\&
\text{TDI}\hspace{3pt} \text{Sagnac}\hspace{3pt} \text{basis}\hspace{3pt} \alpha_{2}\\&
P_{1}=\left(1-\mathcal{D}_{2^{\prime} 1^{\prime} 3^{\prime}}\right), P_{1 \prime}=-\left(1-\mathcal{D}_{312}\right), \\&
P_{2}=\left(1-\mathcal{D}_{2^{\prime} 1^{\prime} 3^{\prime}}\right) \mathcal{D}_{3}, P_{2^{\prime}}=-\left(1-\mathcal{D}_{312}\right) \mathcal{D}_{2^{\prime} 1^{\prime}}, \\&
P_{3}=\left(1-\mathcal{D}_{2^{\prime} 1^{\prime} 3^{\prime}}\right) \mathcal{D}_{31}, P_{3^{\prime}}=-\left(1-\mathcal{D}_{312}\right) \mathcal{D}_{2^{\prime}}
\end{aligned}
\end{equation}

\quad get $\beta_{2}$ and $\gamma_{2}$ though rotation($1\to 2\to 3\to 1$)

\quad The signal-to-noise ratio (SNR) is a crucial parameter for assessing gravitational wave sources. The first-generation TDI was designed to address the equal-arm case, while the second-generation TDI is capable of handling situations where $L_{i}$ and $L_{i^{\prime}}$ differ not only in value but also in their time dependence. It is clear that the first-generation TDI is insufficient in eliminating laser frequency noise of orders higher than speed. This has an impact on the SNRs that are considered significant in contributing to the results.

\quad The purpose of this section is to derive the optimal sensitivity by optimum weighting of second-generation Time Delay Interferometry (TDI) configurations under the following assumptions: 1) Noise independence; 2) Other second-generation TDI configurations can be obtained by linearly combining the generators $\alpha_{2},\beta_{2},\gamma_{2},X_{2}$ in Eq.(\ref{eq:TDIX coefficients}). Tinto has shown that it is possible to derive a family of $\zeta$-like combinations \cite{Tinto2014}\cite{tinto2022TDI2.0}; 3) L1=L2=L3, so D=$D_{i}$ and $D^{2}=D_{ij}$. The generators of $\zeta$-like combinations TDI algebraic space are four.TDI algebraic space can be obtain
\begin{equation}
\centering
\begin{split}
\label{TDI algebraic space}
TDI(f)=\sum\lambda_{i}(f,a)X_{i}
\end{split}
\end{equation}

where $\lambda_{i}(f,a)$ are arbitrary complex functions of the Fourier frequency f, a is Characteristic parameters of gravitational waves,$X_{i}$ are TDI space generator.

\quad SNR can be get by\cite{Tinto2014}
\begin{equation}
\centering
\begin{split}
\label{SNRTDI}
SNR^{2}=\int\frac{|\sum\lambda_{i}(f,a)X^{s}_{i}|^{2}}{|\sum\lambda_{i}(f,a)X^{n}_{i}|^{2}}
\end{split}
\end{equation}

\quad where subscripts s and n refer to the signal and the noise parts of TDI space generator,After a series of algebraic processes such as differentiation and derivation in\cite{Tinto2014}\cite{noise.character}, we can get
\begin{equation}
\centering
\begin{aligned}
\label{TDI optimal}
\mathrm{SNR}_{ \mathrm{opt}}^{2}=\int \mathbf{x}_{i}^{(\mathrm{s}) *}\left(\mathbf{C}^{-1}\right)_{i j} \mathbf{x}_{j}^{(\mathrm{s})}d f
\end{aligned}
\end{equation}
The correlation matrix in TDI generator space is given by $C=< X^{n}_{i}, X^{n}_{j}> $, which is Hermitian and non-singular. Currently, the optimal TDI1.0 configuration, which consists of A, E, and T, has been obtained. By combining these three configurations, the first-generation optimal TDI can increase the SNR value by $\sqrt{2}$ times in the low-frequency range and $\sqrt{3}$ times in the high-frequency range \cite{optimal.LISA.sensitive}.

Based on the previous assumptions, where the matrix C is 4 $\times$ 4,The noise correlation matrix C is uniquely identified by two real functions, Sa and Sab.So the matrix C can be expressed by

\begin{equation}\label{eq:C matrix}
C = \begin{pmatrix}
S_a & S_{ab} & S_{ab} & S_{ab} \\
S_{ab} & S_a & S_{ab} & S_{ab} \\
S_{ab} & S_{ab} & S_a & S_{ab} \\
S_{ab} & S_{ab} & S_{ab} & S_a
\end{pmatrix}
\end{equation}
\quad Based on the previous assumptions, the optimal signal-to-noise ratio can be converted to the sum of the 'converted' signal-to-noise ratio of the four interference combinations. By using Mathematica code, it is easy to obtain the four eigenvalue matrices C.
\begin{equation}
(Sa-Sab, Sa-Sab, Sa-Sab, Sa+3Sa )    
\end{equation}
\quad After orthogonalization of all the eigenvectors, the first three eigenvectors correspond to the same eigenvalue, while the fourth eigenvector corresponds to an eigenvalue orthogonal to them.we get
\begin{equation}
\label{TDI2.0 configuration}
\begin{aligned}
   \\& A_{1}=\frac{1}{\sqrt{2}}  (-\alpha _{2}+ X _{2}) \\&A_{2}=\frac{1}{\sqrt{2}}  (-\alpha _{2}+ \gamma  _{2})
    \\&A_{3}=\frac{1}{\sqrt{2}}  (-\alpha _{2}+\beta  _{2})
     \\&B=\frac{1}{2}  (\alpha
    _{2}+ \beta  _{2}+\gamma   _{2}+ X _{2})
\end{aligned}
\end{equation}

\quad In this section, we obtain the second-generation optimal TDI combination to maximize the SNR value. Although its form is somewhat similar to the first-generation optimal TDI, it has some more interesting properties that will be studied in the following sections.
\section{sensitivity curve }
\subsection{PSD}
\begin{figure}[ht!]
\centering
\includegraphics[width=0.7\columnwidth2]{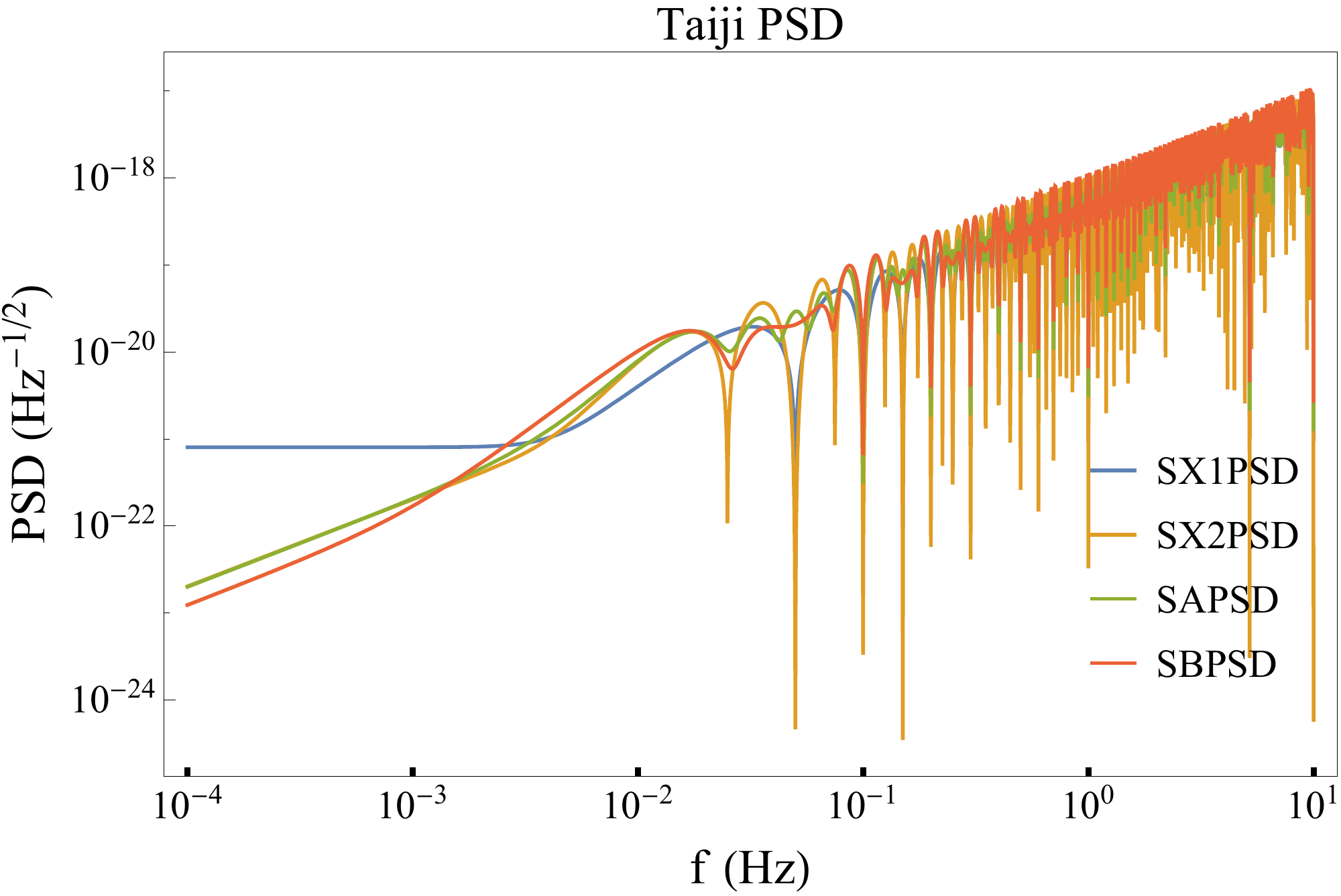}

\caption{The comparison of taiji noise power spectral density (PSD) mainly includes TDI1.0X configuration, TDI2.0X configuration, TDI2.0A configuration and TDI2.0B configuration.}
\label{figs:taiji-PSD}
\end{figure}

\begin{figure}[ht!]
\centering
\includegraphics[width=0.7\columnwidth7]{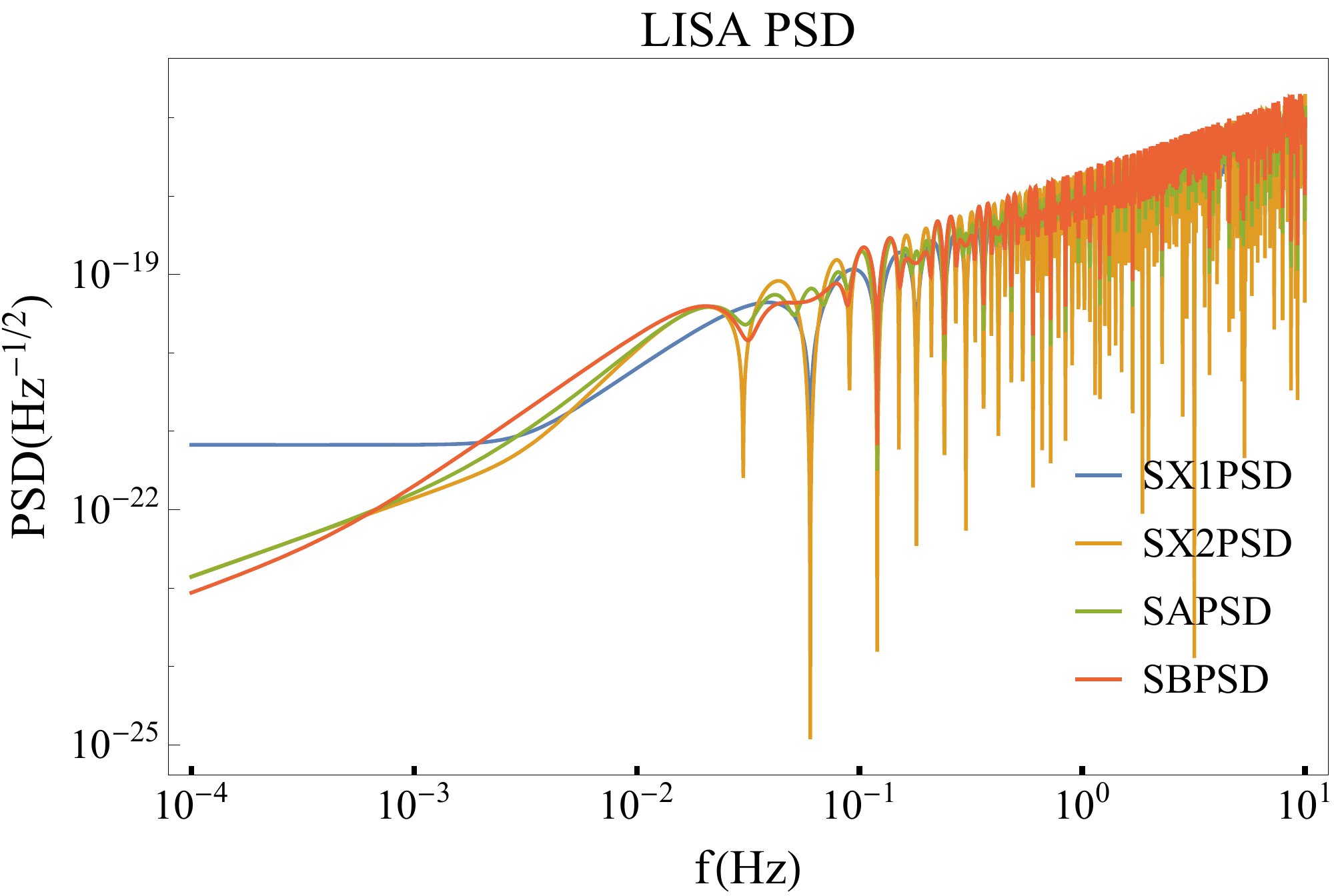}

\caption{The comparison of LISA noise power spectral density (PSD) mainly includes TDI1.0X configuration, TDI2.0X configuration, TDI2.0A configuration and TDI2.0B configuration.}
\label{figs:LISA-PSD}
\end{figure}

\begin{figure}[ht!]
\centering
\includegraphics[width=0.7\columnwidth]{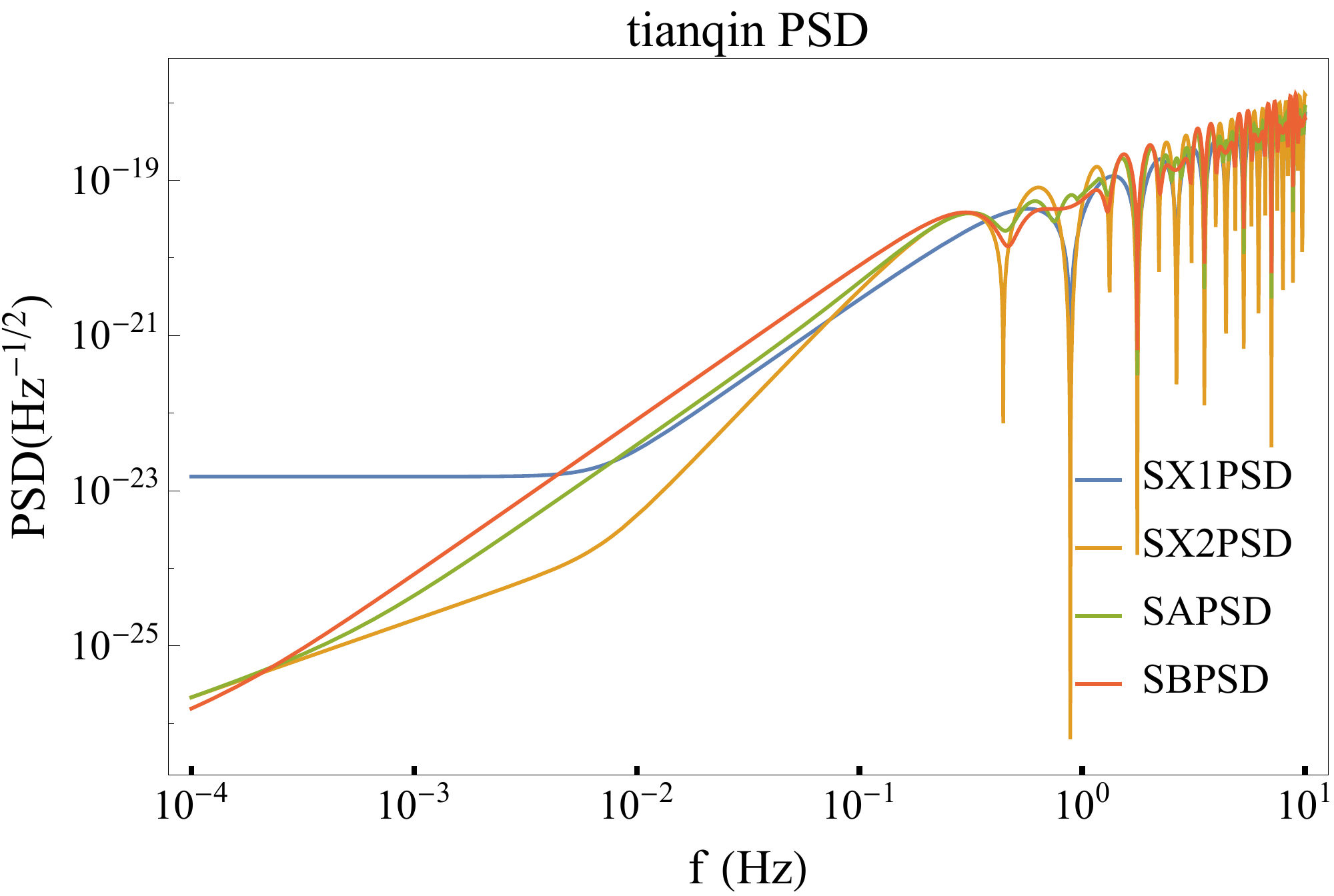}

\caption{The comparison of tianqin noise power spectral density (PSD) mainly includes TDI1.0X configuration, TDI2.0X configuration, TDI2.0A configuration and TDI2.0B configuration.}
\label{figs:Tianqin-PSD}
\end{figure}

\quad To achieve optimal signal-to-noise ratio for the detection of gravitational wave sources, a simple toy model is used that considers the addition of proof mass and shot noise in the noise power spectral density \cite{Karsten.Danzmann_2003,10.1093.nsr.nwx116,Luo_2016}. The resulting linear combination of total residual power spectral densities of proof mass and shot noise can be expressed as \cite{Wangpanpan2021,wangpanpan2022}.
\begin{equation}
\label{eq:PSD}
 \begin{aligned}
PSD(u) & =S_{\operatorname{TDI}^{a}}(u)+S_{\mathrm{TDI}^{\text {shot }}}(u) \\
& =C_{1}\left[\tilde{P}_{i}(u)\right] n_{1}(u)+4 C_{2}\left[\tilde{P}_{i}(u)\right] n_{2}(u),
\end{aligned}
\end{equation}
where $C_{1}$ and $C_{2}$
\begin{equation}
\begin{split}
 \begin{aligned}
\begin{array}{l}
C_{1}\left[\tilde{P}_{i}(u)\right]=\sum_{i=1}^{3} \operatorname{Re}\left[\left|\tilde{P}_{i}\right|^{2}+\left|\tilde{P}_{i^{\prime}}\right|^{2}\right], \\
C_{2}\left[\tilde{P}_{i}(u)\right]=\sum_{i=1}^{3} \operatorname{Re}\left[\tilde{P}_{i} \tilde{P}_{(i+1)^{\prime}}^{*}\right],
\end{array}\\
\end{aligned}
\end{split}
\end{equation}

\begin{equation}
n_{1}(u)=2 S_{\mathrm{pf}}+S_{\mathrm{opt}}, \\
n_{2}(u)=S_{\mathrm{pf}} \cos u,
\end{equation}

with $ u=(2\pi f L/c). S_{pf}=\frac{s_{a}^2}{(2\pi fc)^2} and S_{shot}=  \frac{(2\pi f)^2 s_{x}^2}{c^2}$, where $s_{a}$ and $s_x$ are amplitude spectral densities (ASDs) of proof mass acceleration and shot noises, respectively.
\\
\quad By use In Sec.II TDI coefficient,the PSD are obtained for $L_1=L_2=L_3=L$ in following list:

\begin{equation}
\label{eq:X1PSD}
\begin{aligned}
\operatorname{SX1PSD}[f]=&\left(16(3+\operatorname{Cos}[4 f( L/c) \pi]) \sin [2f L\pi]^{2}\right) S a\\&+Sx 16\operatorname{Sin}[2f(L/c)\pi]^{2}
\end{aligned}
\end{equation}

\begin{equation}
\label{eq:X2PSD}
\begin{aligned}
\operatorname{X2PSD}(f)=&256Sa\operatorname{Cos}[2f(L/c)\pi]^{2}(3+\operatorname{Cos}[4f(L/c)\pi])\\
&\operatorname{Sin}[2 f(L / c) \pi]^{4}+64 \operatorname{Sx} \operatorname{Sin}[2 f(L/c)\pi]^{2}\\
&\operatorname{Sin}[4 f(L/c)\pi]^{2}.
\end{aligned}
\end{equation}

\begin{equation}\label{eq:SAPSD}
\begin{aligned}
\operatorname{SA1PSD}(f)&=\operatorname{SA2PSD}(f)=\operatorname{SA3PSD}(f)\\
&=16 \mathrm{Sa} \sin ^{4}(\pi f(L / c))(344 \cos (2 \pi f(L / c))\\
&+244 \cos (4 \pi f(L / c))+136 \cos (6 \pi f(L / c))\\
&+56 \cos (8 \pi f(L / c))+16 \cos (10 \pi f(L / c))\\
&+4 \cos (12 \pi f(L / c))+195)\\
&+\operatorname{Sx}(4 \cos (2 \pi f(L / c))-2(3 \cos (4 \pi f(L / c))\\
&+4 \cos (6 \pi f(L / c))+4 \cos (8 \pi f(L / c))\\
&+\cos (10 \pi f(L / c))-3 \cos (12 \pi f(L / c))-7))
\end{aligned}
\end{equation}

\begin{equation}\label{eq:XBPSD}
\begin{aligned}
\operatorname{SBPSD}(f)&=8 \operatorname{Sa} \sin ^{4}(\pi f(L / c))(280 \cos (2 \pi f(L / c))\\
&+188 \cos (4 \pi f(L / c))+92 \cos (6 \pi f(L / c))\\
&+28 \cos (8 \pi f(L / c))+4 \cos (10 \pi f(L / c))\\
&+163)+ \mathrm{Sx}(10 \cos (2 \pi f(L / c))-2 \cos (4 \pi f(L / c))\\
&-11 \cos (6 \pi f(L / c))-12 \cos (8 \pi f(L / c))\\
&-3 \cos (10 \pi f(L / c))+2 \cos (12 \pi f(L / c))\\
&+\cos (14 \pi f(L / c))+15)
\end{aligned}
\end{equation}
\quad Reference some published literature parameters for Taiji,LISA,tianqin GW detector in Table.\ref{tab:parameters},The noise comparison diagram is shown in Fig.\ref{figs:taiji-PSD},Fig.\ref{figs:LISA-PSD},Fig.\ref{figs:Tianqin-PSD}.

\begin{table}[htbp]
  \centering
  \caption{Parameters for GW detectors}
    \begin{tabular}{c|c|c|c}
    \hline
    Detector & $L$ ($\mathrm{m}$) & $\mathrm{sa}$ ($\mathrm{m/s^2/\sqrt{Hz}}$) & $\mathrm{sx}$ ($\mathrm{m/\sqrt{Hz}}$) \\
    \hline
    Taiji  & $30\cdot10^{8}$ & $3\cdot10^{-15}$ & $8\cdot10^{-12}$ \\
    LISA   & $25\cdot10^{8}$ & $3\cdot10^{-15}$ & $15\cdot10^{-12}$ \\
    Tianqin & $1.7\cdot10^{8}$ & $1\cdot10^{-15}$ & $1\cdot10^{-12}$ \\
    \hline
    \end{tabular}%
  \label{tab:parameters}%
\end{table}%
\quad where sa is Acceleration noise,sx is shot noise

\subsection{The GW All-sky averaged response function}
\quad In this section, the performance of different second-generation TDI configurations is investigated under the assumption of an average spherical polarization of gravitational waves and a simple noise model. To this end, a published response function \cite{Wangpanpan2021,wangpanpan2022} is used to illustrate the differences between the configurations.

\quad As to common TDI combination reads 
\begin{equation}\label{eq:commo response function}
\begin{aligned}
R(u)&=\frac{2}{4} C_{1}\left[\tilde{P}_{i}(u)\right] \times f_{1}(u)\\&+C_{2}\left[\tilde{P}_{i}(u)\right] \times f_{2}(u)\\&+\frac{3}{4} C_{3}\left[\tilde{P}_{i}(u)\right] \times f_{3}(u)\\&-\frac{3}{4} C_{4}\left[\tilde{P}_{i}(u)\right] \times f_{4}(u)\\&+\frac{1}{4} C_{5}\left[\tilde{P}_{i}(u)\right] \times f_{5}(u)
\end{aligned}
\end{equation}

\begin{equation}
\begin{aligned}
&f_{1}(u)=\frac{4}{3}-\frac{2}{u^{2}}+\frac{\sin 2 u}{u^{3}} \\&
f_{2}(u)=\frac{-u \cos u+\sin u}{u^{3}}-\frac{\cos u}{3} \\&
f_{3}(u)=\log \frac{4}{3}-\frac{5}{18}+\frac{-5 \sin u+8 \sin 2 u-3 \sin 3 u}{8 u} \\&-\frac{1}{3}\left(\frac{4+9 \cos u+12 \cos 2 u+\cos 3 u}{8u^{2}}\right)\\&+\frac{1}{3}\left(\frac{-5 \sin u+8 \sin 2 u+5 \sin 3 u}{8 u^{3}}\right)+\mathrm{Ci} 3 u-2 \mathrm{Ci} 2 u+\mathrm{Ci} u\\&
f_{4}(u)=\frac{-5 \cos u+8 \cos 2 u-3 \cos 3 u}{8 u}-\operatorname{Si} 3 u-\operatorname{Si} u+2 \operatorname{Si} 2 u\\&+\frac{1}{3}\left(\frac{9 \sin u+12 \sin 2 u+\sin 3 u}{8 u^{2}}-\frac{8+5 \cos u}{8 u^{3}}\right)\\&
+\frac{1}{3}\left(-\frac{-8 \cos 2 u-5 \cos 3 u}{8 u^{3}}\right)
\\&
f_{5}(u)=-\log 4+\frac{7}{6}+\frac{11 \sin u-4 \sin 2 u}{4 u}\\&-\frac{10+5 \cos u-2 \cos 2 u}{4 u^{2}}\\&+\frac{5 \sin u+4 \sin 2 u}{4 u^{3}}+2(\mathrm{Ci} 2 u-\mathrm{Ci} u)
\end{aligned}
\end{equation}

\begin{equation}
\begin{aligned}
&C_{1}=\sum_{i=1}^{3}\left[\left.\tilde{P}_{i}\right|^{2}+\left|\tilde{P}_{i^{\prime}}\right|^{2}\right]\\&
C_{2}=2 \sum_{i=1}^{3} \operatorname{Re}\left[\tilde{P}_{i} \tilde{P}_{(i+1)^{\prime}}^{*}\right]\\&
C_{3}=2 \sum_{i=1}^{3} \operatorname{Re}\left[\left(\tilde{P}_{i} \tilde{P}_{i+1}^{*}+\tilde{P}_{i^{\prime}} \tilde{P}_{(i-1)^{\prime}}^{*}\right) e^{i u}\right]\\&
C_{4}=2 \sum_{i=1}^{3} \operatorname{Im}\left[\left(\tilde{P}_{i} \tilde{P}_{i+1}^{*}+\tilde{P}_{i^{\prime}} \tilde{P}_{(i-1)^{\prime}}^{*}\right) e^{i u}\right]\\&
C_{5}=2 \sum_{i=1}^{3} \operatorname{Re}\left[\tilde{P}_{i} \tilde{P}_{i^{\prime}}^{*}+\tilde{P}_{i} \tilde{P}_{(i-1)^{\prime}}^{*}\right]
\end{aligned}
\end{equation}

\begin{figure}[ht!]
\centering
\includegraphics[width=0.7\columnwidth]{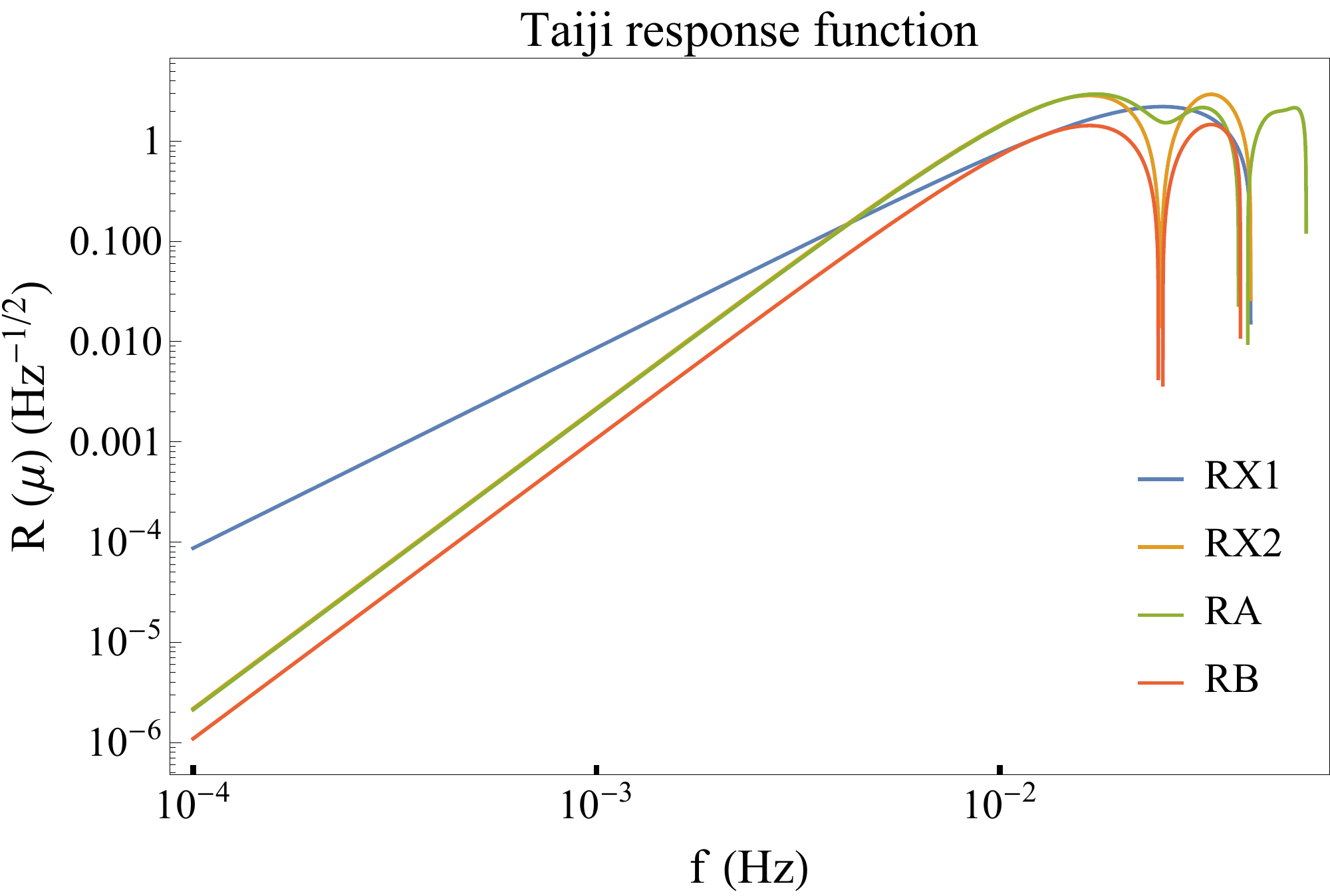}

\caption{The comparison of Taiji GW averaged response function mainly includes TDI1.0X configuration, TDI2.0X configuration, TDI2.0A configuration and TDI2.0B configuration.}
\label{figs:Taiji-response function}
\end{figure}

\begin{figure}[ht!]
\centering
\includegraphics[width=0.7\columnwidth]{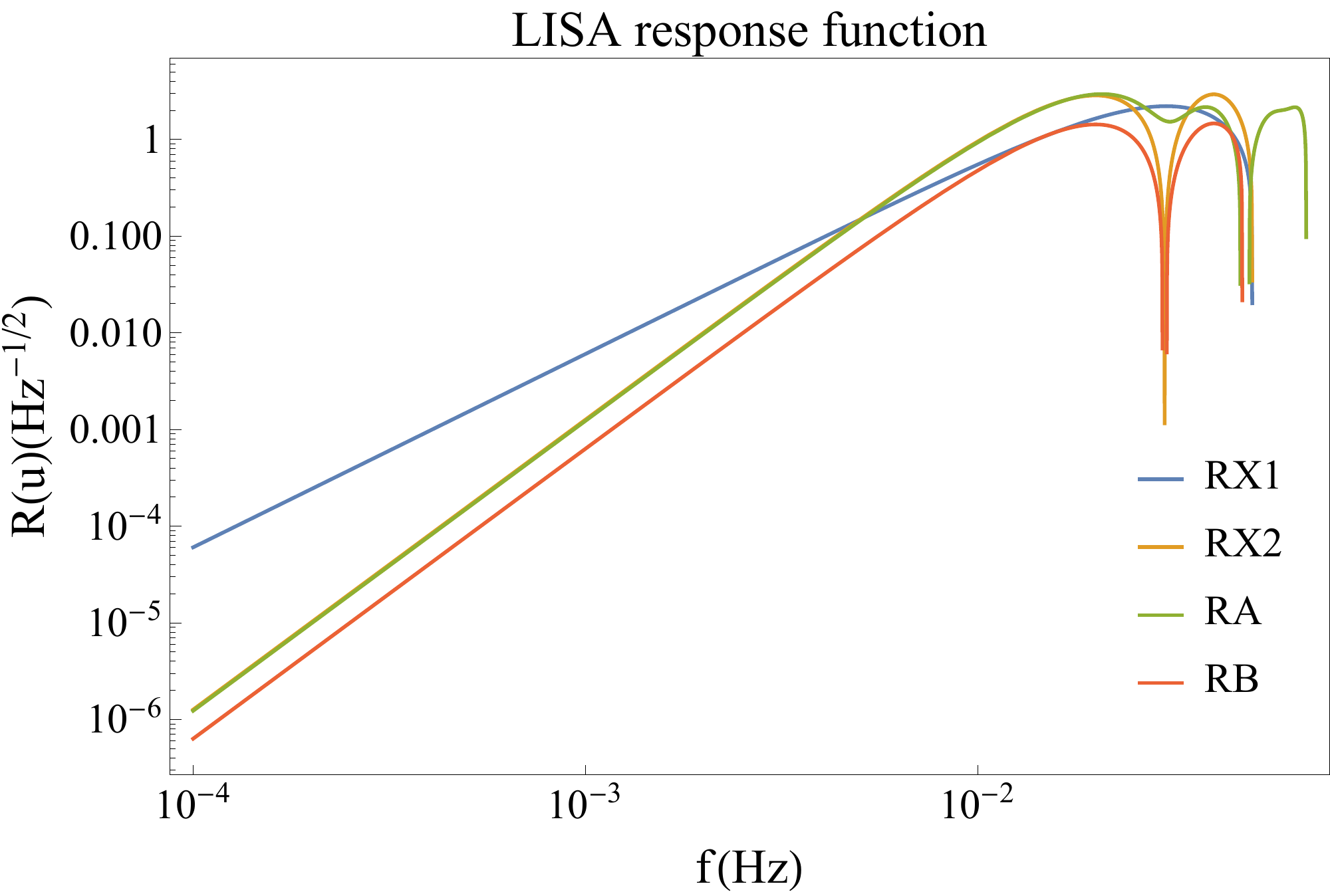}

\caption{The comparison of LISA GW averaged response function mainly includes TDI1.0X configuration, TDI2.0X configuration, TDI2.0A configuration and TDI2.0B configuration.}
\label{figs:LISA-response function}
\end{figure}

\begin{figure}[ht!]
\centering
\includegraphics[width=0.7\columnwidth]{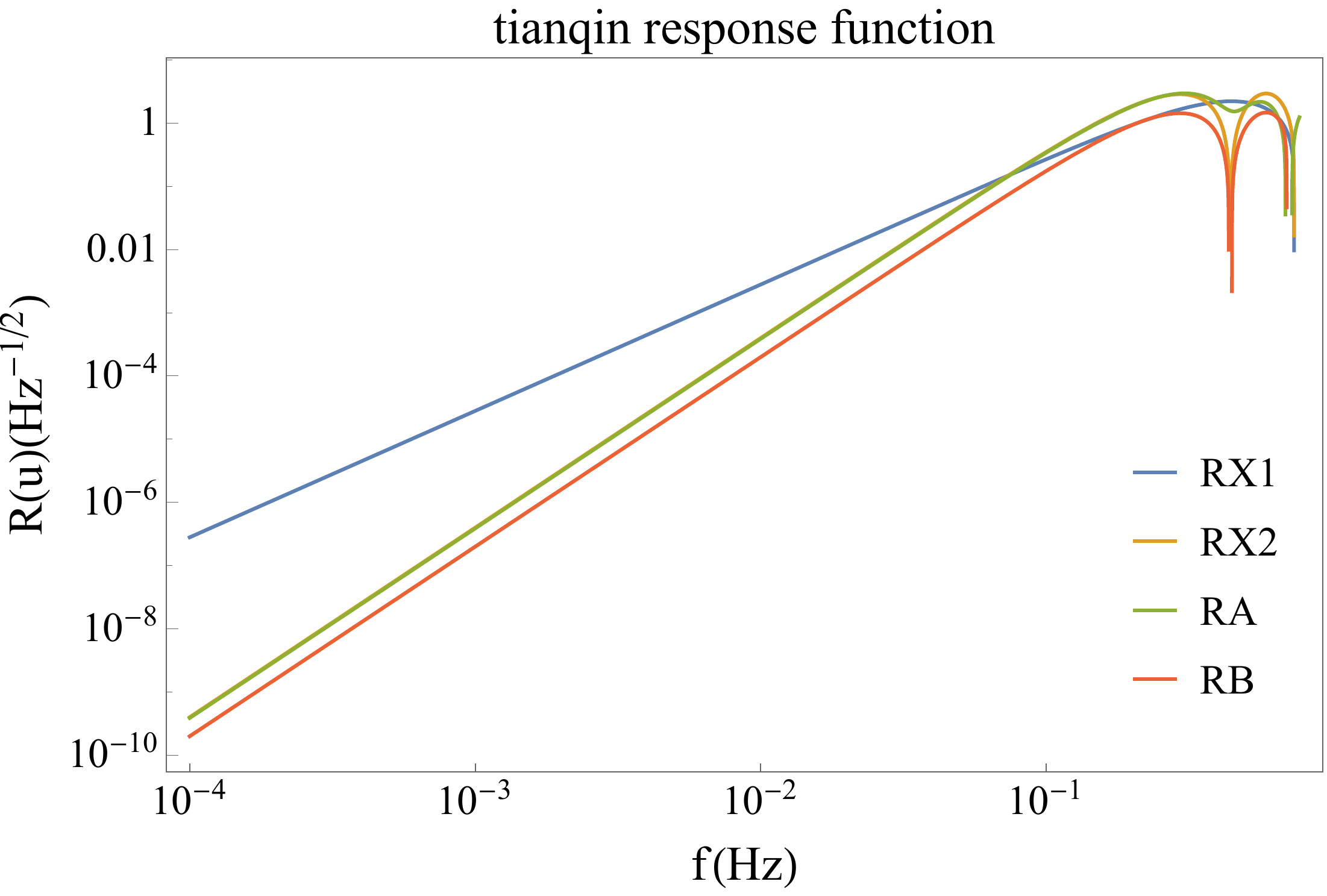}

\caption{The comparison of tianqin GW averaged response function mainly includes TDI1.0X configuration, TDI2.0X configuration, TDI2.0A configuration and TDI2.0B configuration.}
\label{figs:Tianqin-response function}
\end{figure}

For the TDI2.0 A and B combinations, by substitute Eq.\ref{TDI2.0 configuration} into Eq.\ref{eq:commo response function},Detailed calculate results are obtained for needed configuration, and low frequency limit and high frequency limit results can be obtained for similar way. Fig.\ref{figs:Taiji-response function},Fig.\ref{figs:LISA-response function},Fig.\ref{figs:Tianqin-response function} are shown in following list at low frenquence limit.

\subsection{sensitive and Optimization comparison}
The calculation formula of SNR and the construction method of sensitivity curve are briefly introduced.The definition of SNR for all sky average is quoted\cite{Robson2019}:
\begin{equation}
\begin{split}
\begin{array}{l}
SNR^2=T \int \frac{H^2}{PSD(u)/R(u)} df
\end{array}
\end{split}
\end{equation}

\begin{figure}[ht!]
\centering
\includegraphics[width=0.7\columnwidth]{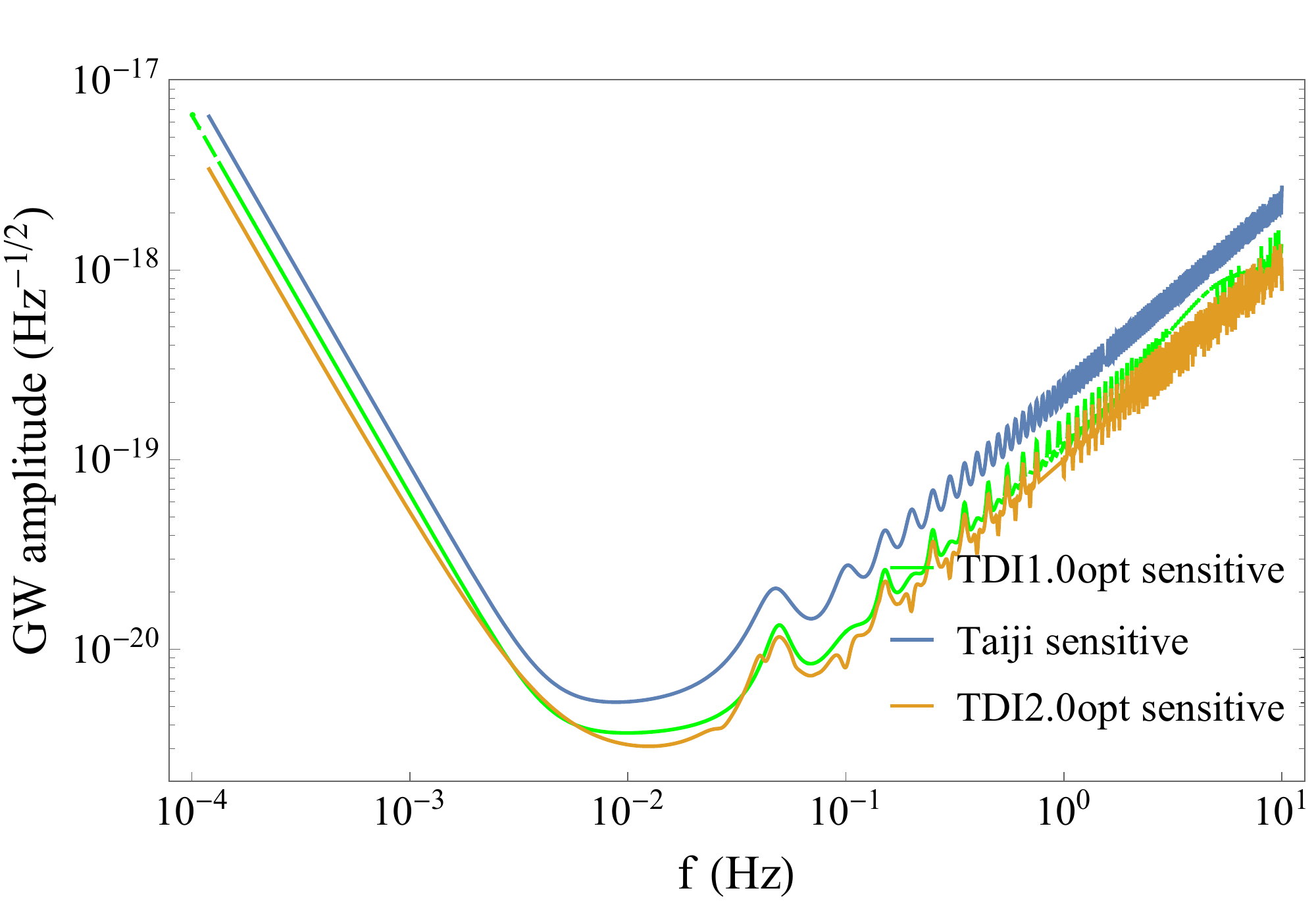}
\caption{TDI2.0opt is superior to the current best TDI1.0opt, which has a certain enhancement effect for the detection of dense binary wave source bands, while the effect is extremely significant for the SMBH wave source bands}
\label{figs:TDI.0 optimal and TDI2.0 optimal compare}
\end{figure}

\quad where H is GW amplitude in frequency domain and $PSD(u)$ and $R(u)$ are derive in equation(\ref{eq:PSD})(\ref{eq:commo response function}).

\quad In order to align with the existing literature\cite{Robson2019},$T=1$and$SNR=1$,the sensitive form read as.
\begin{equation}
\begin{split}
\begin{array}{l}
Sensitive(u)=\sqrt{{PSD(u)/R(u)}} 
\end{array}
\end{split}
\end{equation}

\begin{figure}[ht!]
\centering
\includegraphics[width=0.7\columnwidth]{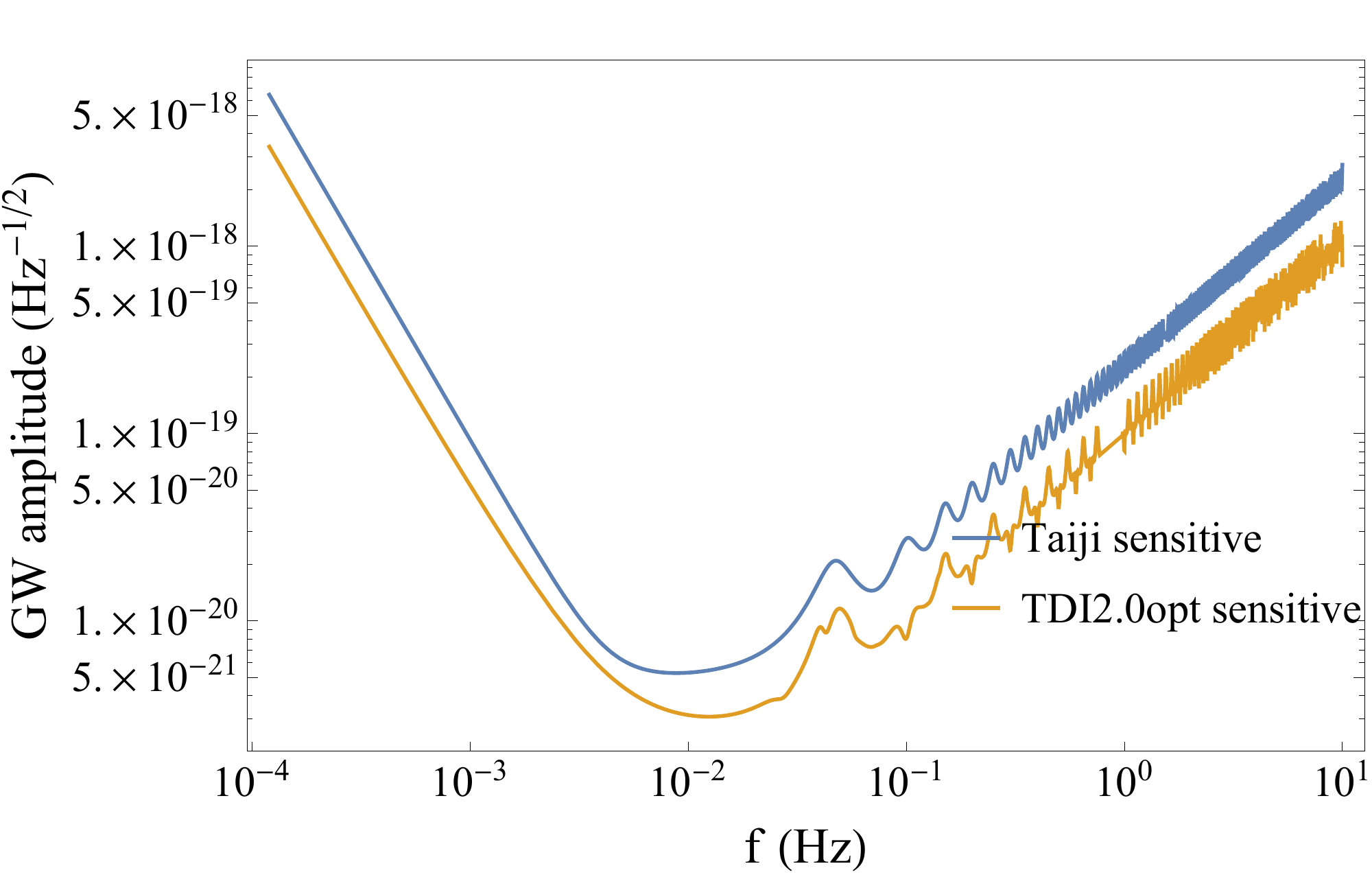}
\hspace{1in}
\includegraphics[width=0.7\columnwidth]{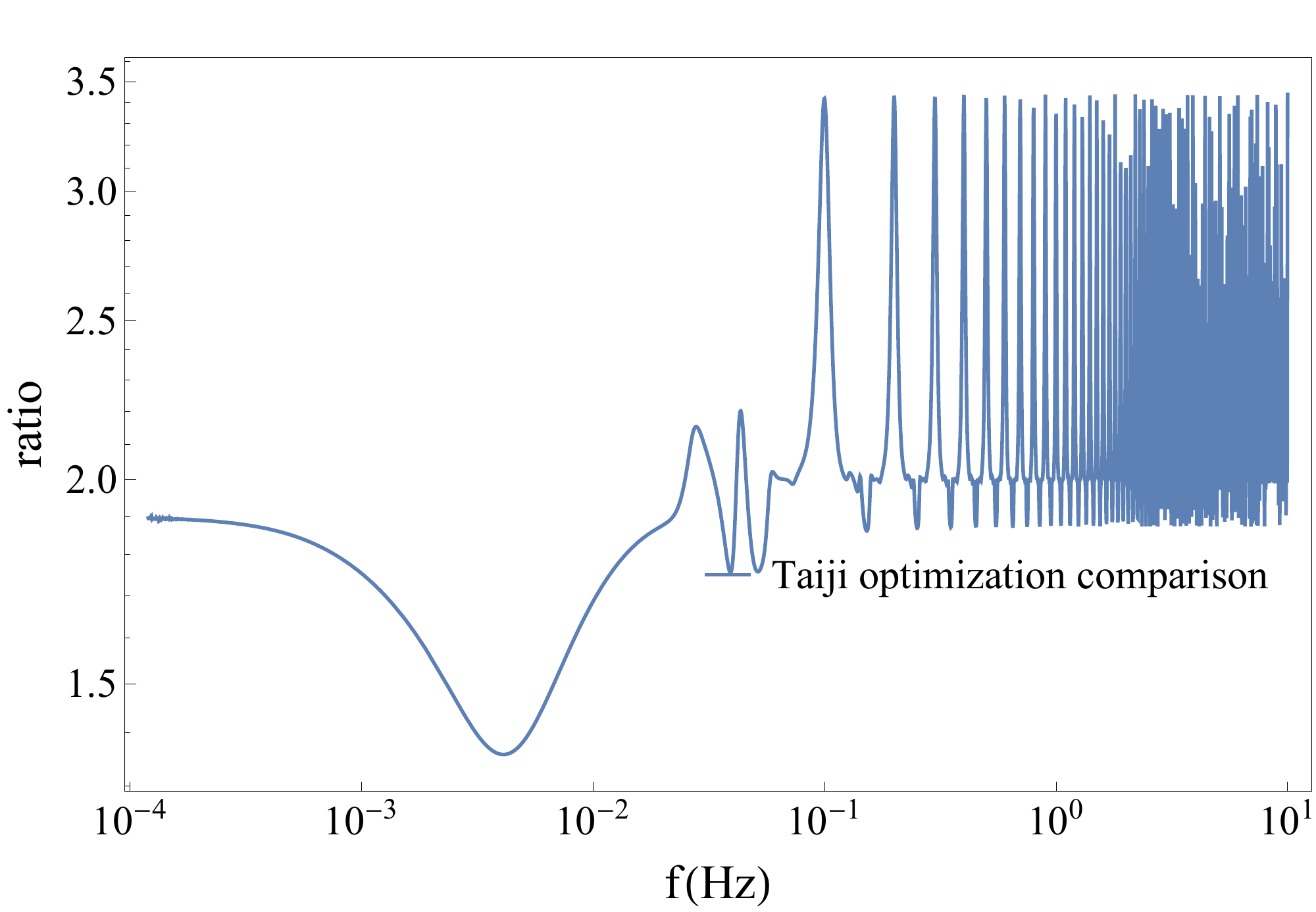}
\caption{The figure above shows the configuration sensitivity curves of TDI2.0opt and TDIX with taiji parameter, and the figure below shows the relative optimization efficiency of TDI2.0opt}
\label{figs:taiji sensitive and optimization comparison}
\end{figure}

\begin{figure}[htbp]
\centering
\includegraphics[width=0.7\columnwidth]{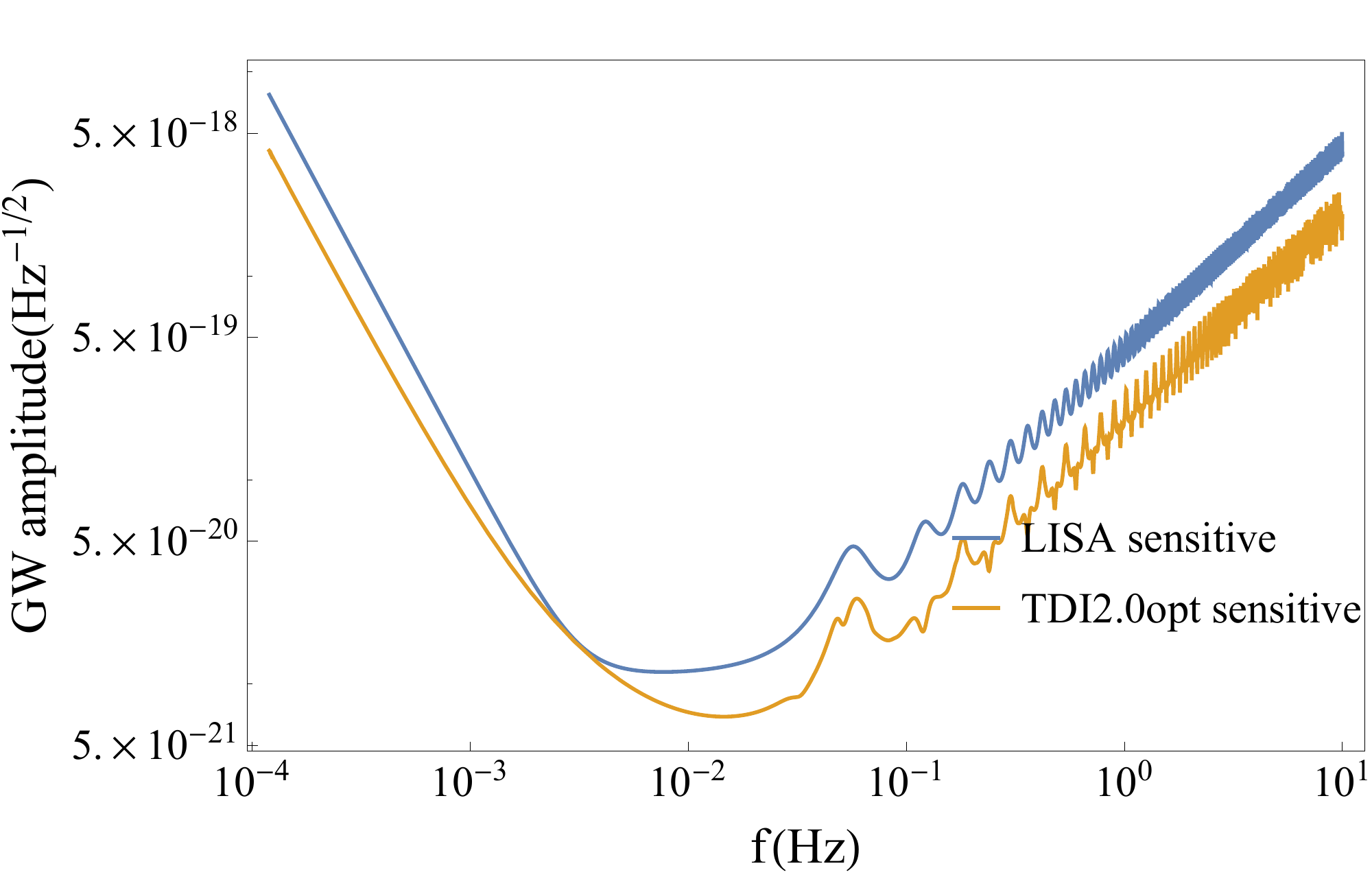}
\hspace{1in}
\includegraphics[width=0.7\columnwidth]{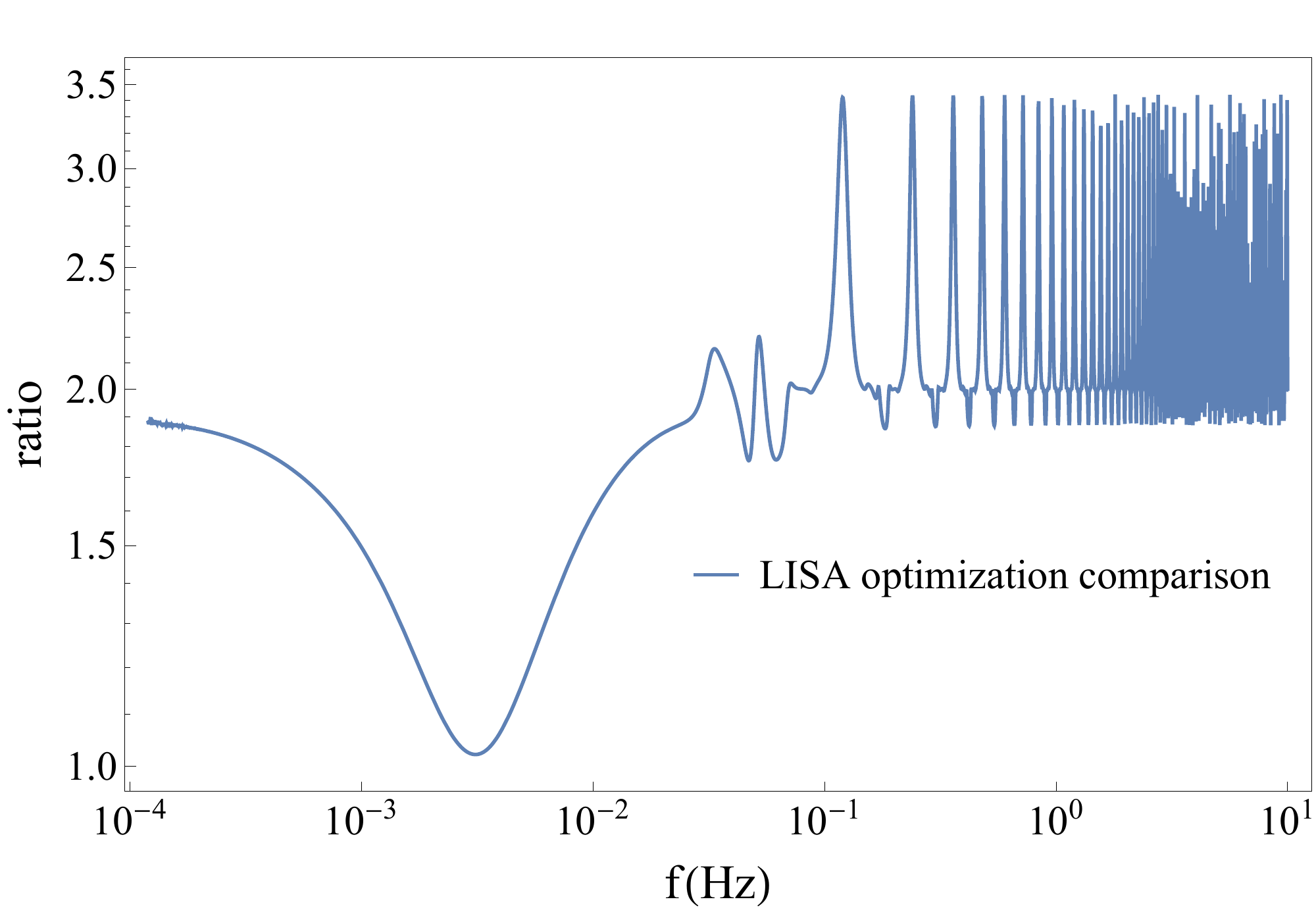}
\caption{The figure above shows the configuration sensitivity curves of TDI2.0opt and TDIX with LISA parameter, and the figure below shows the relative optimization efficiency of TDI2.0opt}
\label{figs: LISA sensitive and optimization comparison}
\end{figure}

\begin{figure}[t!]
\centering
\includegraphics[width=0.7\columnwidth]{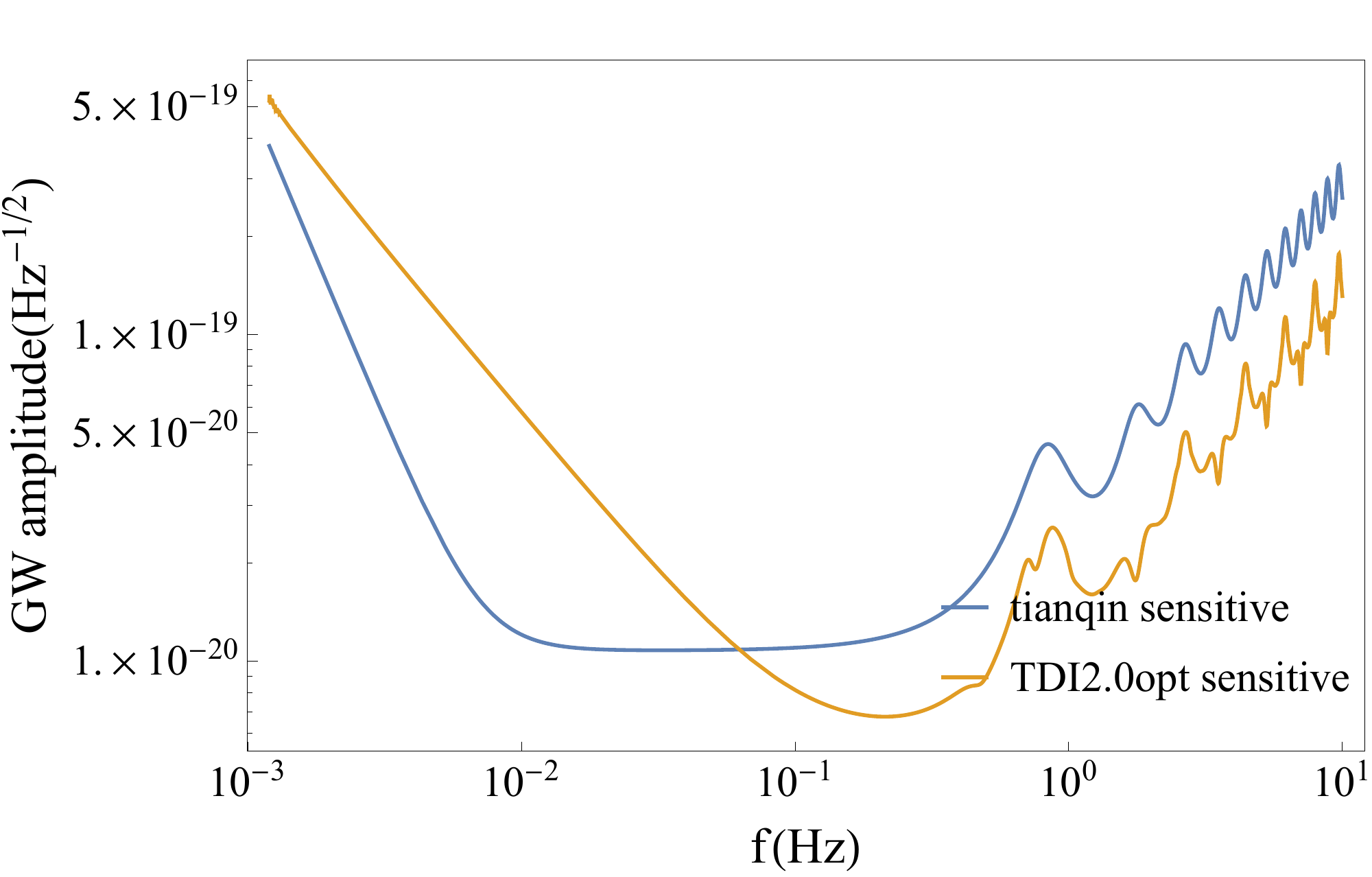}
\hspace{1in}
\includegraphics[width=0.6\columnwidth]{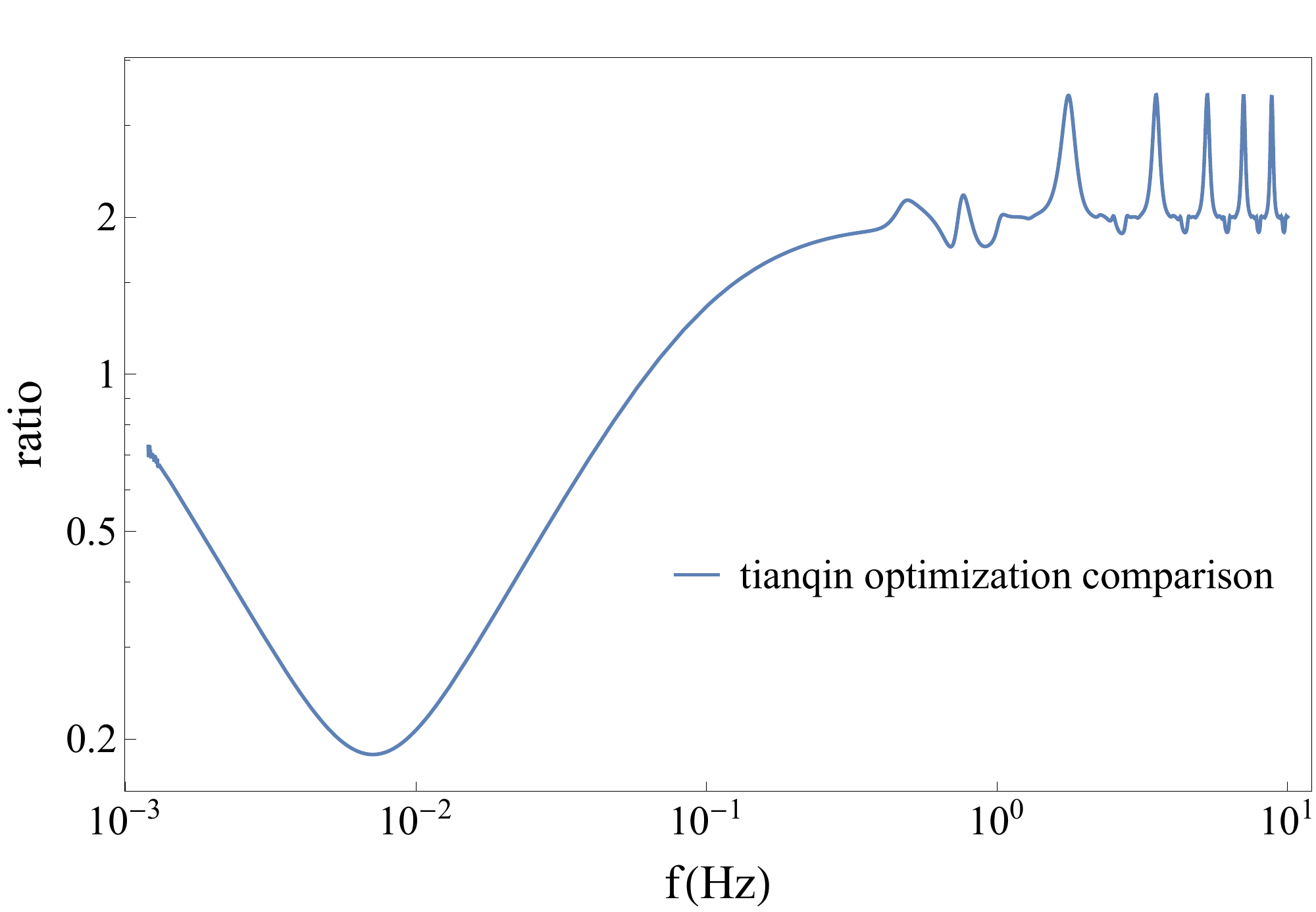}

\caption{The figure above shows the configuration sensitivity curves of TDI2.0opt and TDIX with tianqin parameter, and the figure below shows the relative optimization efficiency of TDI2.0opt}
\label{figs:tianqin sensitive and optimization comparison}
\end{figure}

\quad We show sensitivity curves for LISA, Taiji, and Tianqin detectors, and compare them with the most sensitive X configuration. Additionally, we compare the sensitivity of first-generation and second-generation optimal TDI for Taiji parameters. It is noteworthy that the X1 configuration of the first-generation TDI has the same sensitivity as the X2 configuration of the second-generation TDI. (Figures 
\ref{figs: LISA sensitive and optimization comparison}, 
\ref{figs:taiji sensitive and optimization comparison}, \ref{figs:tianqin sensitive and optimization comparison}, and \ref{figs:TDI.0 optimal and TDI2.0 optimal compare}).

\subsection{Explore the factors influencing optimal TDI2.0}

\quad In this part, only three influential factors such as arm length, shot noise and accelerate noise were considered, and explore the change of sensitivity curve by change the parameter value.
\begin{figure}[htbp!]
\centering
\includegraphics[width=0.7\columnwidth]{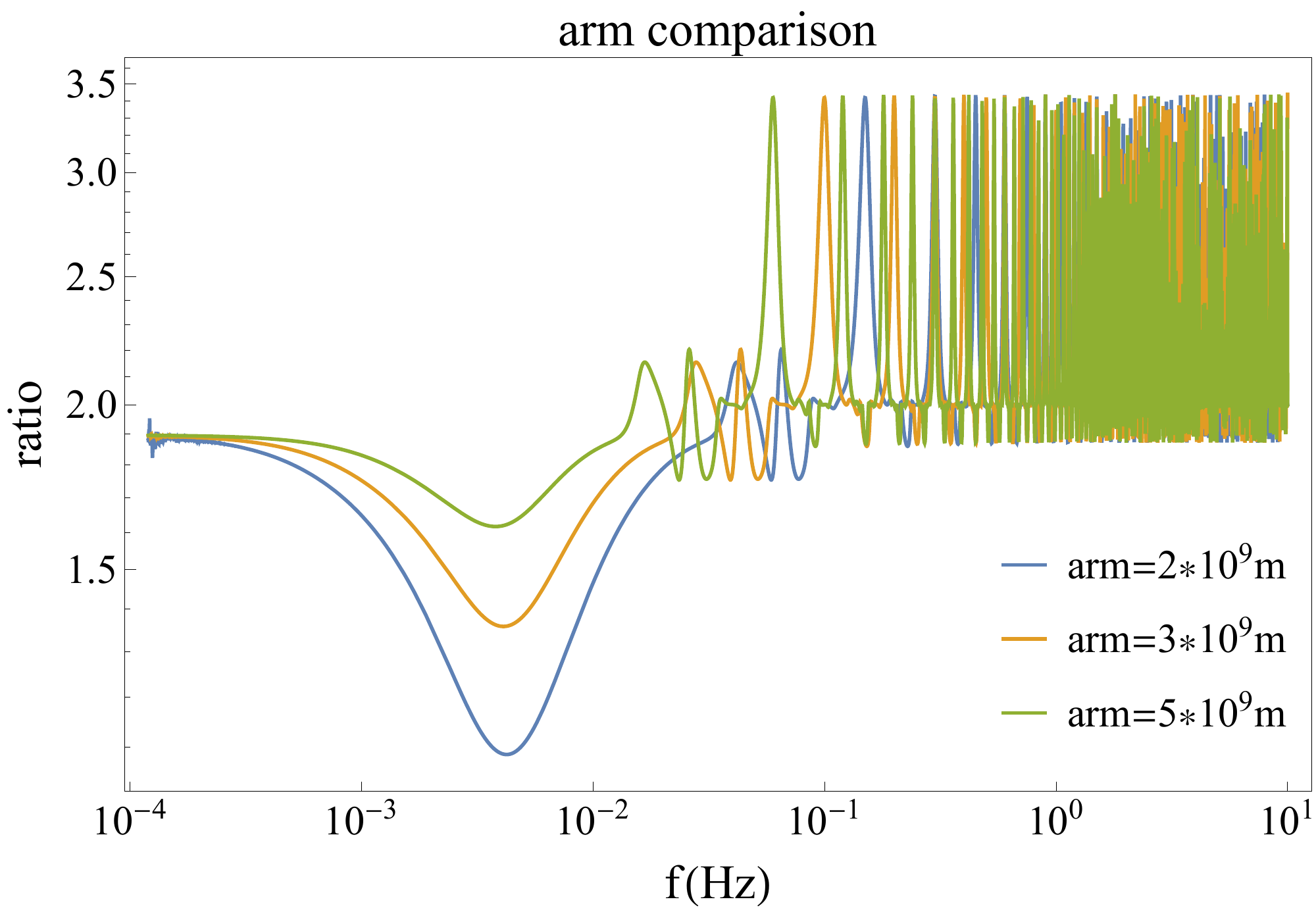}
\hspace{1in}
\includegraphics[width=0.7\columnwidth]{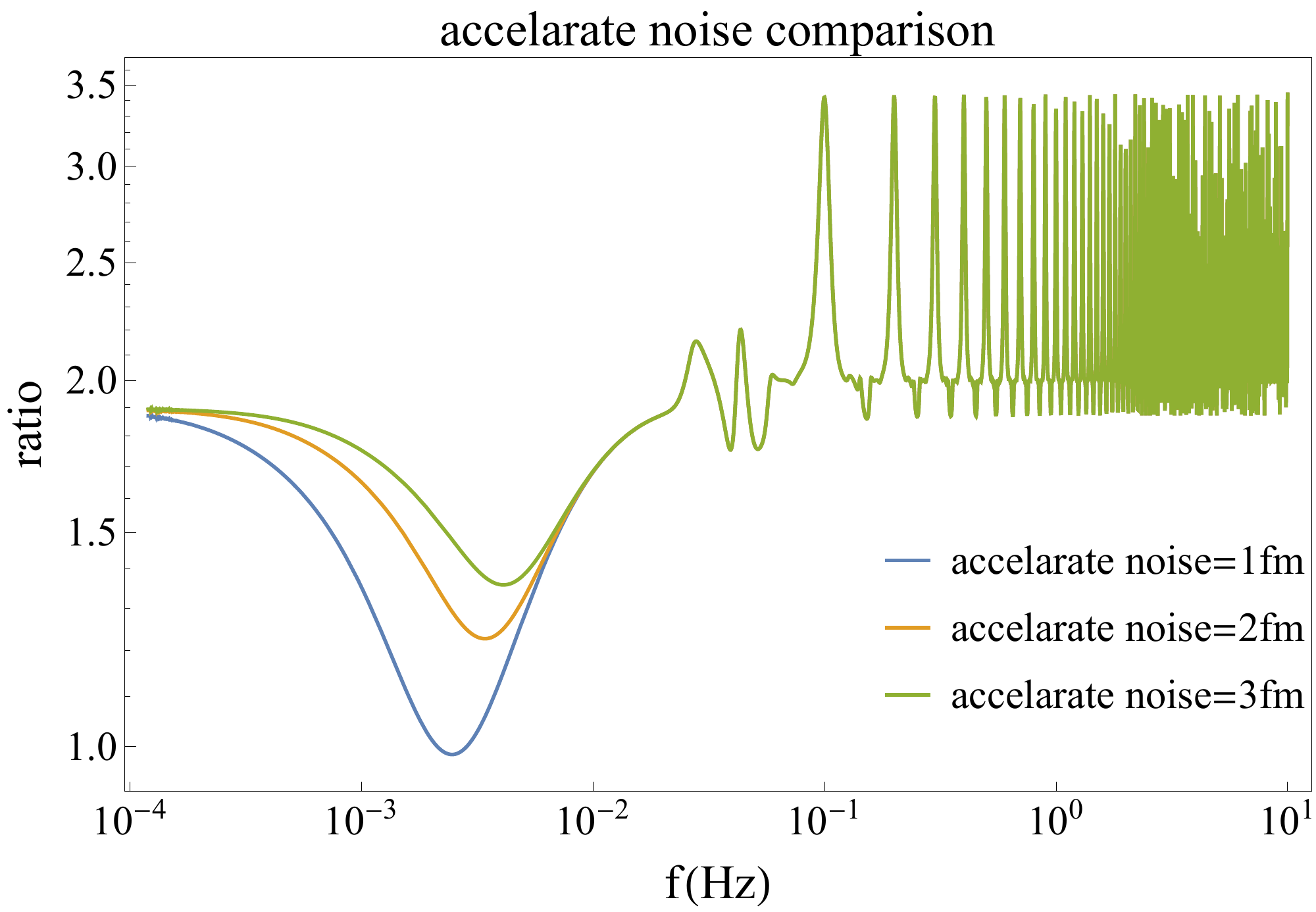}
\hspace{1in}
\includegraphics[width=0.7\columnwidth]{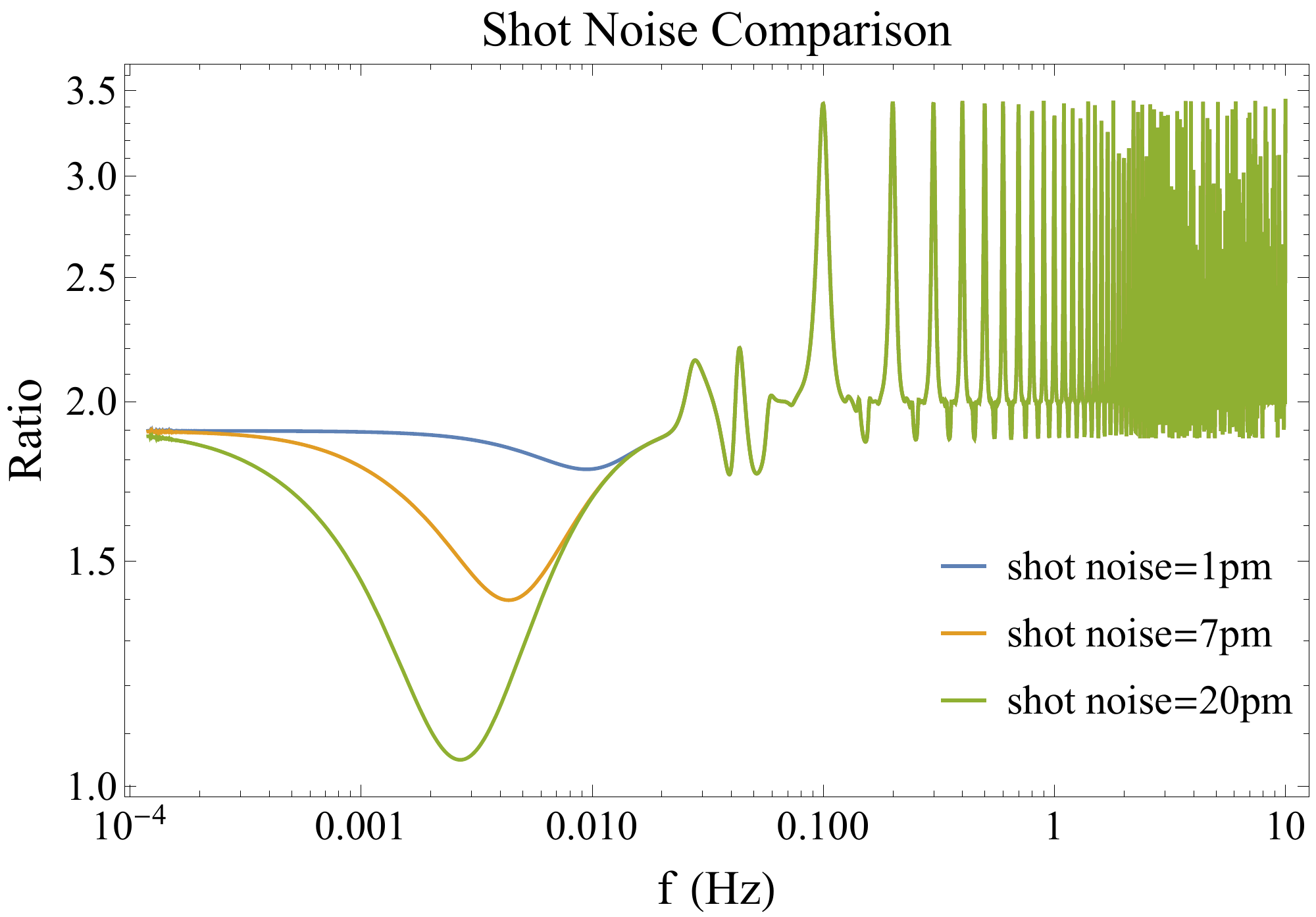}
\caption{The analysis focuses on taiji parameters, while other parameters yield similar results. Each comparison varies only the target parameter while keeping other parameters constant. The results indicate that the three factor parameters have no impact on high frequencies. In the low frequency region, arm length has a linear effect, with TDI2.0opt yielding better optimization with longer arm lengths. Shot noise suppression improves the optimization effect, while acceleration noise suppression reduces it.}
\label{figs:factor influence optimal TDI2.0 comparison}
\end{figure}

\section{SNR values and detection rates}

\subsection{Double White Dwarf}
\quad White dwarfs are highly compact objects located except to neutron stars and black holes. According to the existing cosmological and white dwarf formation models, it is estimated that there are about $10^{9}$ double white dwarf systems \cite{DWB.verification.source}. Although there are many theoretical models of white dwarf formation mechanisms, such as CO+CO, CO+He, He+He, etc. \cite{DWD.pupulation}, the evolution time of double white dwarfs is usually millions of years \cite{WDBGravitationalWaveSignal2001}\cite{WDB.forget}, which is far longer than the duration of current space-based gravitational wave observation missions. As a result, we can calculate the SNR value at a single frequency point, as the double white dwarfs evolve slowly in the frequency domain. The main consideration in this case is the detection rate of white dwarfs. For a continuous gravitational wave signal radiated by a compact double white dwarf system \cite{WDBGravitationalWaveSignal2001}, its amplitude can be approximated as follows.
\begin{equation}
\label{eq:hWD1}
\begin{aligned}
h(n, e)&=\left[\frac{16 \pi G}{c^{3} \omega_{\mathrm{g}}^{2}} \frac{L(n, e)}{4 \pi d^{2}}\right]^{1 / 2} \\&
=1.010^{-21} \frac{\sqrt{g(n, e)}}{n}\left(\frac{\mathcal{M}}{\mathrm{M}_{\odot}}\right)^{5 / 3}\left(\frac{P_{\mathrm{orb}}}{1 \mathrm{hr}}\right)^{-2 / 3}\left(\frac{d}{1 \mathrm{kpc}}\right)^{-1}\\
g(n, e)&=\frac{1+(73 / 24) e^{2}+(37 / 96) e^{4}}{\left(1-e^{2}\right)^{7 / 2}}
\end{aligned}
\end{equation}

If we don't think about the higher harmonic term and choose Approximate circular orbit,e=0,n=2
\begin{equation}
\label{eq:hWD2}
\begin{aligned}
h(0,2)&=5.05 \times 10^{-22}\left(\frac{\mathcal{M}}{\mathrm{M}_{\odot}}\right)^{5 / 3}\left(\frac{P_{\mathrm{orb}}}{1 \mathrm{hr}}\right)^{-2 / 3}\left(\frac{d}{1 \mathrm{kpc}}\right)^{-1}
\end{aligned}
\end{equation}

\quad Introduce relation between $f_{gw}$ and $P_{orb}$  of DWD binary mass which mass is \cite{TDI1.0opt.direction}.

\begin{equation}
\label{eq:fgw and prob}
\begin{aligned}
f_{g w}=\frac{2}{P_{o r b}}=2.3\left(\frac{P_{o r b}}{.01 \mathrm{day}}\right)^{-1} \mathrm{mHz}
\end{aligned}
\end{equation}

\quad substitude Equation (\ref{eq:fgw and prob}) into Equation (\ref{eq:hWD2})

\begin{equation}
\label{eq:fgw(0,0) and prob}
\begin{aligned}
h(0,2)&=5.05 \times 10^{-22}\left(\frac{\mathcal{M}}{\mathrm{M}_{\odot}}\right)^{5 / 3}
\\& \left(\frac{(2.3\cdot 10^{-3}/f_{gw})\cdot 0.01\cdot 24hr}{1 \mathrm{hr}}\right)^{-2 / 3}\left(\frac{d}{1 \mathrm{kpc}}\right)^{-1}.
\end{aligned}
\end{equation}
\quad Taking into account the all sky average condition,And we get a concrete expression for SNR

\begin{equation}
\begin{split}
\label{eq:WDBSNR}
\begin{array}{l}
SNR^2=5 T \frac{(\frac{4}{5})^2 \widetilde{h}(f)^2}{PSD(u)/R(u)} 
\end{array}
\end{split}
\end{equation}
\quad Using the Taiji parameters, we calculate the all-sky average and verification sources of White Dwarf sources\cite{DWB.verification.source} for different TDI configurations using the above formula. The detailed results are shown in Appendix A.To explore the detection rate of White Dwarf sources(SNR $>$ 8) under different TDI optimals , we randomly sample 1 million White Dwarf sources from the following parameter region using a random function.
\begin{equation}
\begin{aligned}
&m_{1}\sim [0.1,1]M_{\odot}\\
&m_{2}\sim [0.01,0.1]M_{\odot}\\
&D_{eff}\sim [10,50]Mpc\\
&f\sim [0.001,0.01]Hz
\end{aligned}
\end{equation}

\begin{figure}[H]
\centering
\includegraphics[scale=0.34]{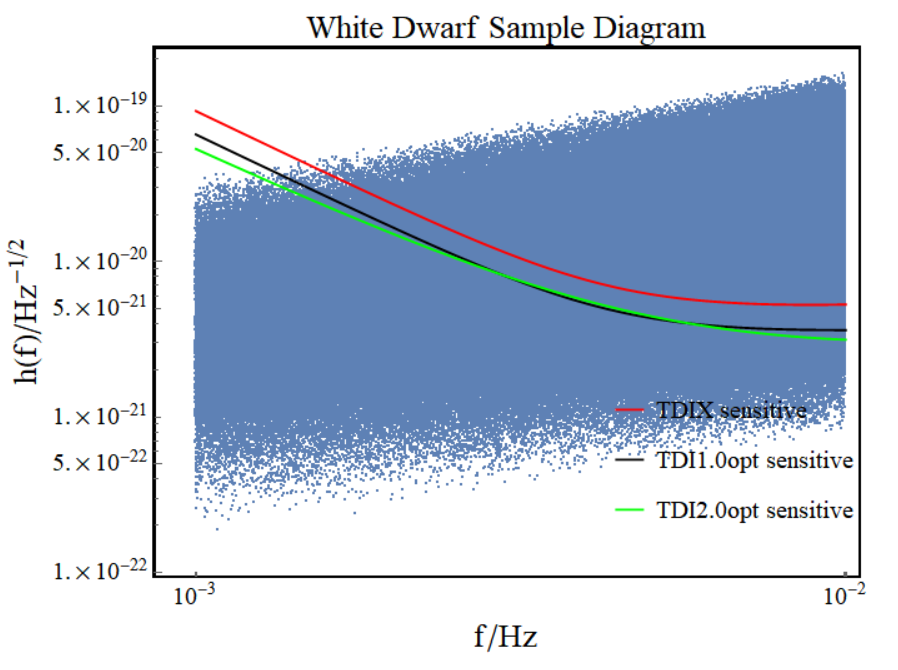}

\caption{One million samples of White Dwarf binaries (DWBS) were generated. With a threshold of SNR$>$8, 53088 events were detected using TDIx, resulting in a detection rate of 5.30$\%$. Similarly, 111950 events were detected using TDI1.0 optimal, resulting in a detection rate of 11.20$\%$, while TDI2.0 optimal detected 130568 events with a detection rate of 13.05$\%$.}
\label{WDB sample diagram}
\end{figure}

\quad During code calculate(Fig.\ref{WDB sample diagram}),The threshold is SNR$>$8, TDIx get 53088 event and detection rate is 5.30$\%$;TDI1.0 optimal get 111950 event and detection rate is 11.20$\%$;TDI2.0 optimal get 130568 event and detection rate is 
13.05$\%$.

\newpage
\section{Conclusion}
\quad In this study, we have presented the fundamental principles and applications of the second-generation optimal Time Delay Interferometry (TDI) technique. We began by providing a concise overview of the first-generation optimal TDI process, followed by an explanation of the conceptual foundation and development trajectory of the second-generation optimal TDI.
\quad Subsequently, we derived the power spectral density formula and sensitivity curve of the second-generation optimal TDI, comparing them to other TDI configurations. In the final section, we investigated the impact of various detector parameters on the optimization performance of TDI 2.0. Moreover, we assessed the signal-to-noise ratio and detection rate of White Dwarf systems based on the optimal TDI configuration.
\quad Based on this comprehensive analysis, we will consider different configuration effect on global fit problem and parameter estimate accuracy in future .

\bibliographystyle{IEEEtran}
\bibliography{TDI_references}

\begin{thebibliography}{10}
\providecommand{\url}[1]{#1}
\csname url@samestyle\endcsname
\providecommand{\newblock}{\relax}
\providecommand{\bibinfo}[2]{#2}
\providecommand{\BIBentrySTDinterwordspacing}{\spaceskip=0pt\relax}
\providecommand{\BIBentryALTinterwordstretchfactor}{4}
\providecommand{\BIBentryALTinterwordspacing}{\spaceskip=\fontdimen2\font plus
\BIBentryALTinterwordstretchfactor\fontdimen3\font minus
  \fontdimen4\font\relax}
\providecommand{\BIBforeignlanguage}[2]{{%
\expandafter\ifx\csname l@#1\endcsname\relax
\typeout{** WARNING: IEEEtran.bst: No hyphenation pattern has been}%
\typeout{** loaded for the language `#1'. Using the pattern for}%
\typeout{** the default language instead.}%
\else
\language=\csname l@#1\endcsname
\fi
#2}}
\providecommand{\BIBdecl}{\relax}
\BIBdecl

\bibitem{abbottObservationGravitationalWaves2016}
\BIBentryALTinterwordspacing
B.~P. Abbott, R.~Abbott, T.~D. Abbott, M.~R. Abernathy, F.~Acernese, K.~Ackley,
  C.~Adams, and Adams, ``Observation of gravitational waves from a binary black
  hole merger,'' \emph{Phys. Rev. Lett.}, vol. 116, p. 061102, Feb 2016.
  [Online]. Available:
  \url{https://link.aps.org/doi/10.1103/PhysRevLett.116.061102}
\BIBentrySTDinterwordspacing

\bibitem{PhysRevLett.116.061102}
\BIBentryALTinterwordspacing
B.~P. Abbott, R.~Abbott, T.~D. Abbott, M.~R. Abernathy, F.~Acernese, and
  K.~Ackley, ``Observation of gravitational waves from a binary black hole
  merger,'' \emph{Phys. Rev. Lett.}, vol. 116, p. 061102, Feb 2016. [Online].
  Available: \url{https://link.aps.org/doi/10.1103/PhysRevLett.116.061102}
\BIBentrySTDinterwordspacing

\bibitem{PhysRevD.93.021101}
\BIBentryALTinterwordspacing
W.~Chaibi, R.~Geiger, B.~Canuel, A.~Bertoldi, A.~Landragin, and P.~Bouyer,
  ``Low frequency gravitational wave detection with ground-based atom
  interferometer arrays,'' \emph{Phys. Rev. D}, vol.~93, p. 021101, Jan 2016.
  [Online]. Available:
  \url{https://link.aps.org/doi/10.1103/PhysRevD.93.021101}
\BIBentrySTDinterwordspacing

\bibitem{Accadia_2011}
\BIBentryALTinterwordspacing
T.~Accadia, F.~Acernese, F.~Antonucci, P.~Astone, G.~Ballardin, F.~Barone,
  M.~Barsuglia, A.~Basti, T.~S. Bauer, M.~Bebronne, M.~G. Beker, A.~Belletoile,
  S.~Birindelli, M.~Bitossi, and M.~A. Bizouard, ``Status of the virgo
  project,'' \emph{Classical and Quantum Gravity}, vol.~28, no.~11, p. 114002,
  may 2011. [Online]. Available:
  \url{https://dx.doi.org/10.1088/0264-9381/28/11/114002}
\BIBentrySTDinterwordspacing

\bibitem{Harry_2010}
\BIBentryALTinterwordspacing
G.~M. Harry and (forthe LIGO Scientific~Collaboration), ``Advanced ligo: the
  next generation of gravitational wave detectors,'' \emph{Classical and
  Quantum Gravity}, vol.~27, no.~8, p. 084006, apr 2010. [Online]. Available:
  \url{https://dx.doi.org/10.1088/0264-9381/27/8/084006}
\BIBentrySTDinterwordspacing

\bibitem{Abbott_2009}
\BIBentryALTinterwordspacing
B.~P. Abbott, R.~Abbott, R.~Adhikari, P.~Ajith, B.~Allen, G.~Allen, and R.~S.
  Amin, ``Ligo: the laser interferometer gravitational-wave observatory,''
  \emph{Reports on Progress in Physics}, vol.~72, no.~7, p. 076901, jun 2009.
  [Online]. Available: \url{https://dx.doi.org/10.1088/0034-4885/72/7/076901}
\BIBentrySTDinterwordspacing

\bibitem{Kawamura_2011}
\BIBentryALTinterwordspacing
S.~Kawamura, M.~Ando, N.~Seto, S.~Sato, T.~Nakamura, and K.~Tsubono, ``The
  japanese space gravitational wave antenna: Decigo,'' \emph{Classical and
  Quantum Gravity}, vol.~28, no.~9, p. 094011, apr 2011. [Online]. Available:
  \url{https://dx.doi.org/10.1088/0264-9381/28/9/094011}
\BIBentrySTDinterwordspacing

\bibitem{Belczynski_2010}
\BIBentryALTinterwordspacing
K.~Belczynski, M.~Benacquista, and T.~Bulik, ``Double compact objects as
  low-frequency gravitational wave sources,'' \emph{The Astrophysical Journal},
  vol. 725, no.~1, p. 816, nov 2010. [Online]. Available:
  \url{https://dx.doi.org/10.1088/0004-637X/725/1/816}
\BIBentrySTDinterwordspacing

\bibitem{Rodriguez_2006}
\BIBentryALTinterwordspacing
C.~Rodriguez, G.~B. Taylor, R.~T. Zavala, A.~B. Peck, L.~K. Pollack, and R.~W.
  Romani, ``A compact supermassive binary black hole system,'' \emph{The
  Astrophysical Journal}, vol. 646, no.~1, p.~49, jul 2006. [Online].
  Available: \url{https://dx.doi.org/10.1086/504825}
\BIBentrySTDinterwordspacing

\bibitem{Karsten.Danzmann_2003}
\BIBentryALTinterwordspacing
K.~Danzmann and A.~Rüdiger, ``Lisa technology—concept, status, prospects,''
  \emph{Classical and Quantum Gravity}, vol.~20, no.~10, p.~S1, apr 2003.
  [Online]. Available: \url{https://dx.doi.org/10.1088/0264-9381/20/10/301}
\BIBentrySTDinterwordspacing

\bibitem{10.1093.nsr.nwx116}
\BIBentryALTinterwordspacing
W.-R. Hu and Y.-L. Wu, ``The taiji program in space for gravitational wave
  physics and the nature of gravity,'' \emph{National Science Review}, vol.~4,
  no.~5, pp. 685--686, 2017-10. [Online]. Available:
  \url{https://doi.org/10.1093/nsr/nwx116}
\BIBentrySTDinterwordspacing

\bibitem{Luo_2016}
\BIBentryALTinterwordspacing
J.~Luo, L.-S. Chen, H.-Z. Duan, Y.-G. Gong, S.~Hu, J.~Ji, Q.~Liu, J.~Mei,
  V.~Milyukov, M.~Sazhin, C.-G. Shao, V.~T. Toth, H.-B. Tu, Y.~Wang, Y.~Wang,
  H.-C. Yeh, M.-S. Zhan, Y.~Zhang, V.~Zharov, and Z.-B. Zhou, ``Tianqin: a
  space-borne gravitational wave detector,'' \emph{Classical and Quantum
  Gravity}, vol.~33, no.~3, p. 035010, jan 2016. [Online]. Available:
  \url{https://dx.doi.org/10.1088/0264-9381/33/3/035010}
\BIBentrySTDinterwordspacing

\bibitem{PhysRevD.60.082001}
\BIBentryALTinterwordspacing
K.~S. Thorne and C.~J. Winstein, ``Human gravity-gradient noise in
  interferometric gravitational-wave detectors,'' \emph{Phys. Rev. D}, vol.~60,
  p. 082001, Sep 1999. [Online]. Available:
  \url{https://link.aps.org/doi/10.1103/PhysRevD.60.082001}
\BIBentrySTDinterwordspacing

\bibitem{PhysRevD.70.081101}
\BIBentryALTinterwordspacing
D.~A. Shaddock, B.~Ware, R.~E. Spero, and M.~Vallisneri, ``Postprocessed
  time-delay interferometry for lisa,'' \emph{Phys. Rev. D}, vol.~70, p.
  081101, Oct 2004. [Online]. Available:
  \url{https://link.aps.org/doi/10.1103/PhysRevD.70.081101}
\BIBentrySTDinterwordspacing

\bibitem{Otto2015}
\BIBentryALTinterwordspacing
M.~Otto, ``Time-delay interferometry simulations for the laser interferometer
  space antenna,'' \emph{Ph.D. thesis, Gottfried Wilhelm Leibniz Universität
  Hannover}, Dec 2015. [Online]. Available: \url{https://doi.org/10.15488/8545}
\BIBentrySTDinterwordspacing

\bibitem{AET99}
\BIBentryALTinterwordspacing
J.~W. Armstrong, F.~B. Estabrook, and M.~Tinto, ``Time-delay interferometry for
  space-based gravitational wave searches,'' \emph{The Astrophysical Journal},
  vol. 527, no.~2, pp. 814--826, dec 1999. [Online]. Available:
  \url{https://doi.org/10.1086/2F308110}
\BIBentrySTDinterwordspacing

\bibitem{amaro-seoaneLaserInterferometerSpace2017}
\BIBentryALTinterwordspacing
P.~Amaro-Seoane, H.~Audley, S.~Babak, and Baker. Laser interferometer space
  antenna. [Online]. Available: \url{http://arxiv.org/abs/1702.00786}
\BIBentrySTDinterwordspacing

\bibitem{babakLISASensitivitySNR2021}
\BIBentryALTinterwordspacing
S.~Babak, M.~Hewitson, and A.~Petiteau. Lisa sensitivity and snr calculations.
  [Online]. Available: \url{http://arxiv.org/abs/2108.01167}
\BIBentrySTDinterwordspacing

\bibitem{GeometricTDI}
\BIBentryALTinterwordspacing
M.~Vallisneri, ``Geometric time delay interferometry,'' \emph{Phys. Rev. D},
  vol.~72, p. 042003, Aug 2005. [Online]. Available:
  \url{https://link.aps.org/doi/10.1103/PhysRevD.72.042003}
\BIBentrySTDinterwordspacing

\bibitem{TDI1.0}
\BIBentryALTinterwordspacing
M.~Tinto and J.~W. Armstrong, ``Cancellation of laser noise in an unequal-arm
  interferometer detector of gravitational radiation,'' \emph{Phys. Rev. D},
  vol.~59, p. 102003, Apr 1999. [Online]. Available:
  \url{https://link.aps.org/doi/10.1103/PhysRevD.59.102003}
\BIBentrySTDinterwordspacing

\bibitem{TDI2.0}
\BIBentryALTinterwordspacing
D.~A. Shaddock, M.~Tinto, F.~B. Estabrook, and J.~W. Armstrong, ``Data
  combinations accounting for lisa spacecraft motion,'' \emph{Phys. Rev. D},
  vol.~68, p. 061303, Sep 2003. [Online]. Available:
  \url{https://link.aps.org/doi/10.1103/PhysRevD.68.061303}
\BIBentrySTDinterwordspacing

\bibitem{TDI1.0opt}
\BIBentryALTinterwordspacing
K.~R. Nayak, A.~Pai, S.~V. Dhurandhar, and J.-Y. Vinet, ``Improving the
  sensitivity of lisa,'' \emph{Classical and Quantum Gravity}, vol.~20, no.~7,
  p. 1217, mar 2003. [Online]. Available:
  \url{https://dx.doi.org/10.1088/0264-9381/20/7/301}
\BIBentrySTDinterwordspacing

\bibitem{Algebraic.approach}
\BIBentryALTinterwordspacing
S.~V. Dhurandhar, K.~R. Nayak, and J.-Y. Vinet, ``Algebraic approach to
  time-delay data analysis for lisa,'' \emph{Phys. Rev. D}, vol.~65, p. 102002,
  May 2002. [Online]. Available:
  \url{https://link.aps.org/doi/10.1103/PhysRevD.65.102002}
\BIBentrySTDinterwordspacing

\bibitem{TDI1.0opt.direction}
\BIBentryALTinterwordspacing
K.~R. Nayak, S.~Dhurandhar, A.~Pai, and J.-Y. Vinet, ``Erratum: Optimizing the
  directional sensitivity of lisa [phys. rev. d 68, 122001 (2003)],''
  \emph{Phys. Rev. D}, vol.~70, p. 049901, Aug 2004. [Online]. Available:
  \url{https://link.aps.org/doi/10.1103/PhysRevD.70.049901}
\BIBentrySTDinterwordspacing

\bibitem{optimal.LISA.sensitive}
\BIBentryALTinterwordspacing
K.~R. Nayak, A.~Pai, S.~V. Dhurandhar, and J.-Y. Vinet, ``Improving the
  sensitivity of lisa,'' \emph{Classical and Quantum Gravity}, vol.~20, no.~7,
  p. 1217, mar 2003. [Online]. Available:
  \url{https://dx.doi.org/10.1088/0264-9381/20/7/301}
\BIBentrySTDinterwordspacing

\bibitem{TDIdatastream}
\BIBentryALTinterwordspacing
F.~B. Estabrook, M.~Tinto, and J.~W. Armstrong, ``Time-delay analysis of lisa
  gravitational wave data: Elimination of spacecraft motion effects,''
  \emph{Phys. Rev. D}, vol.~62, p. 042002, Jul 2000. [Online]. Available:
  \url{https://link.aps.org/doi/10.1103/PhysRevD.62.042002}
\BIBentrySTDinterwordspacing

\bibitem{Tinto2014}
\BIBentryALTinterwordspacing
M.~Tinto and S.~Dhurandhar, ``Time-delay interferometry,'' \emph{Living Reviews
  in Relativity}, vol.~17, 08 2014. [Online]. Available:
  \url{https://doi.org/10.12942/lrr-2014-6}
\BIBentrySTDinterwordspacing

\bibitem{Wangpanpan2021}
\BIBentryALTinterwordspacing
P.-P. Wang, Y.-J. Tan, W.-L. Qian, and C.-G. Shao, ``Sensitivity functions of
  space-borne gravitational wave detectors for arbitrary time-delay
  interferometry combinations regarding nontensorial polarizations,''
  \emph{Phys. Rev. D}, vol. 104, p. 023002, Jul 2021. [Online]. Available:
  \url{https://link.aps.org/doi/10.1103/PhysRevD.104.023002}
\BIBentrySTDinterwordspacing

\bibitem{tinto2022TDI2.0}
\BIBentryALTinterwordspacing
M.~Tinto, S.~Dhurandhar, and D.~Malakar, ``Second-generation time-delay
  interferometry,'' 2022. [Online]. Available:
  \url{https://doi.org/10.48550/arXiv.2212.05967}
\BIBentrySTDinterwordspacing

\bibitem{noise.character}
\BIBentryALTinterwordspacing
J.~Sylvestre and M.~Tinto, ``Noise characterization for lisa,'' \emph{Phys.
  Rev. D}, vol.~68, p. 102002, Nov 2003. [Online]. Available:
  \url{https://link.aps.org/doi/10.1103/PhysRevD.68.102002}
\BIBentrySTDinterwordspacing

\bibitem{wangpanpan2022}
\BIBentryALTinterwordspacing
P.-P. Wang, W.-L. Qian, Y.-J. Tan, H.-Z. Wu, and C.-G. Shao, ``Geometric
  approach for the modified second generation time delay interferometry,''
  \emph{Phys. Rev. D}, vol. 106, p. 024003, Jul 2022. [Online]. Available:
  \url{https://link.aps.org/doi/10.1103/PhysRevD.106.024003}
\BIBentrySTDinterwordspacing

\bibitem{Robson2019}
\BIBentryALTinterwordspacing
T.~Robson, N.~J. Cornish, and C.~Liu, ``The construction and use of lisa
  sensitivity curves,'' \emph{Classical and Quantum Gravity}, vol.~36, no.~10,
  p. 105011, apr 2019. [Online]. Available:
  \url{https://dx.doi.org/10.1088/1361-6382/ab1101}
\BIBentrySTDinterwordspacing

\bibitem{DWB.verification.source}
\BIBentryALTinterwordspacing
V.~Korol, E.~M. Rossi, P.~J. Groot, G.~Nelemans, S.~Toonen, and A.~G.~A. Brown,
  ``Prospects for detection of detached double white dwarf binaries with gaia,
  lsst and lisa,'' \emph{Monthly Notices of the Royal Astronomical Society},
  vol. 470, no.~2, pp. 1894--1910, 2017. [Online]. Available:
  \url{https://doi.org/10.1093/mnras/stx1285}
\BIBentrySTDinterwordspacing

\bibitem{DWD.pupulation}
\BIBentryALTinterwordspacing
{Korol, V.}, {Toonen, S.}, {Klein, A.}, {Belokurov, V.}, {Vincenzo, F.},
  {Buscicchio, R.}, {Gerosa, D.}, {Moore, C. J.}, {Roebber, E.}, {Rossi, E.
  M.}, and {Vecchio, A.}, ``Populations of double white dwarfs in milky way
  satellites and their detectability with lisa,'' \emph{A\&A}, vol. 638, p.
  A153, 2020. [Online]. Available:
  \url{https://doi.org/10.1051/0004-6361/202037764}
\BIBentrySTDinterwordspacing

\bibitem{WDBGravitationalWaveSignal2001}
\BIBentryALTinterwordspacing
G.~Nelemans, L.~R. Yungelson, and S.~F.~P. Zwart, ``The gravitational wave
  signal from the galactic disk population of binaries containing two compact
  objects,'' \emph{A\&A}, vol. 375, no.~3, pp. 890--898, 2001-09-01. [Online].
  Available:
  \url{https://www.aanda.org/articles/aa/abs/2001/33/aah2754/aah2754.html}
\BIBentrySTDinterwordspacing

\bibitem{WDB.forget}
\BIBentryALTinterwordspacing
H.~Li, T.~T. Zhang, T.~Yilmaz, Y.~Y. Pai, C.~E. Marvinney, A.~Said, Q.~W. Yin,
  C.~S. Gong, Z.~J. Tu, E.~Vescovo, C.~S. Nelson, R.~G. Moore, S.~Murakami,
  H.~C. Lei, H.~N. Lee, B.~J. Lawrie, and H.~Miao, ``Observation of
  unconventional charge density wave without acoustic phonon anomaly in kagome
  superconductors ${A\mathrm{V}}_{3}{\mathrm{sb}}_{5}$ ($a=\mathrm{Rb}$, cs),''
  \emph{Phys. Rev. X}, vol.~11, p. 031050, Sep 2021. [Online]. Available:
  \url{https://link.aps.org/doi/10.1103/PhysRevX.11.031050}
\BIBentrySTDinterwordspacing

\end{thebibliography}

\section{Appendix}
\subsection{Appendix  A}
\label{verification WDB}
\begin{table}[!ht]
\centering
\begin{threeparttable}
\caption{SNR of verification WDB by different  TDI optimal}
    \centering
    \begin{tabular}{|l|l|l|l|l|l|l|l|}
        \hline
        source & m1 & m2 & D & f & TDIX & TDI1.0 opt & TDI2.0 opt  \\ \hline
        RX J0806 & 0.55 & 0.27 & 5000.0 & 6.22 & 73.20466546 & 103.2185783 & 106.9031338  \\ 
        V407 Vul  & 0.6 & 0.07 & 2000.0 & 3.51 & 23.35760705 & 32.93422594 & 32.06067987  \\ 
        ES Cet a & 0.6 & 0.06 & 1000.0 & 3.22 & 33.47431685 & 47.19878676 & 46.58364111  \\ 
        AM CVn  & 0.71 & 0.13 & 600.0 & 1.94 & 38.19802502 & 53.85921527 & 59.84772028  \\ 
        SDSS J1908 +3940 & 0.6 & 0.05 & 1000.0 & 1.83 & 6.971176208 & 9.829358453 & 11.06409805  \\ 
        HP Lib & 0.57 & 0.06 & 200.0 & 1.81 & 39.01688777 & 55.01381175 & 62.07102615  \\ 
        PTF1J1919+4815 & 0.6 & 0.04 & 2000.0 & 1.48 & 1.604789843 & 2.262753678 & 2.655675388  \\ 
        CR Boo & 0.79 & 0.06 & 340.0 & 1.36 & 13.56142402 & 19.12160786 & 22.76736853  \\ 
        KL Dra & 0.6 & 0.02 & 1000.0 & 1.33 & 1.222345144 & 1.723506653 & 2.059498637  \\ 
        V803 Cen & 0.84 & 0.08 & 350.0 & 1.25 & 14.54524331 & 20.50879307 & 24.74228728  \\ 
        SDSS J0926a & 0.85 & 0.04 & 460.0 & 1.18 & 4.858215239 & 6.850083488 & 8.333222696  \\ 
        CP Eri & 0.6 & 0.02 & 700.0 & 1.18 & 1.27107637 & 1.792217682 & 2.180257962  \\ 
        2003aw & 0.6 & 0.02 & 700.0 & 0.99 & 0.796850852 & 1.123559701 & 1.397711946  \\ 
        2QZ 1427 -01 & 0.6 & 0.015 & 700.0 & 0.91 & 0.478818637 & 0.675134277 & 0.847658674  \\ 
        SDSS J1240 & 0.6 & 0.01 & 400.0 & 0.89 & 0.527949775 & 0.744409182 & 0.936774109  \\ 
        SDSS J0804 & 0.6 & 0.01 & 400.0 & 0.75 & 0.334608933 & 0.471798595 & 0.603077458  \\ 
        SDSS J1411 & 0.6 & 0.01 & 400.0 & 0.72 & 0.300111961 & 0.423157865 & 0.542663268  \\ 
        GP Com & 0.6 & 0.01 & 80.0 & 0.72 & 1.500559803 & 2.115789323 & 2.713316342  \\ 
        SDSS J0902 & 0.6 & 0.01 & 500.0 & 0.69 & 0.214340172 & 0.302219643 & 0.388814525  \\ 
        SDSS J1552 & 0.6 & 0.01 & 500.0 & 0.59 & 0.14119636 & 0.199086868 & 0.25877857  \\ 
        CE 315 & 0.6 & 0.006 & 77.0 & 0.51 & 0.373830458 & 0.527100945 & 0.690411278  \\ 
        J0651+2844 & 0.55 & 0.25 & 1000.0 & 2.61 & 73.4804528 & 103.6074384 & 107.1526116  \\ 
        J0935+4411 & 0.32 & 0.14 & 660.0 & 1.68 & 14.15056171 & 19.95229201 & 22.86245848  \\ 
        J0106-1000 & 0.43 & 0.17 & 2400.0 & 0.85 & 0.953669274 & 1.344673677 & 1.699849017  \\ 
        J1630+ 4233 & 0.31 & 0.52 & 830.0 & 0.84 & 5.288218978 & 7.456388759 & 9.436507768  \\ 
        J1053+ 5200 & 0.2 & 0.26 & 1100.0 & 0.54 & 0.482579313 & 0.680436831 & 0.888758012  \\ 
        J0923+ 3028 & 0.279 & 0.37 & 228.0 & 0.51 & 3.538424739 & 4.989178882 & 6.534963372  \\ 
        J1436 + 50107 & 0.24 & 0.46 & 800.0 & 0.51 & 1.051639414 & 1.482811574 & 1.942227278  \\ 
        WD 0957-666 & 0.32 & 0.37 & 135.0 & 0.38 & 3.064335551 & 4.320713126 & 5.721886223  \\ 
        J0755+ 4906 & 0.176 & 0.81 & 2620.0 & 0.37 & 0.157200549 & 0.221652774 & 0.293754057  \\ 
        J0849+ 0445 & 0.176 & 0.65 & 1004.0 & 0.29 & 0.182363771 & 0.257132917 & 0.342640677  \\ 
        J0022-1014 & 0.21 & 0.375 & 1151.0 & 0.29 & 0.122846528 & 0.173213604 & 0.230814581  \\ 
        J2119-0018 & 0.74 & 0.158 & 2610.0 & 0.27 & 0.057628499 & 0.081256184 & 0.108406889  \\ 
        J1234-0228 & 0.09 & 0.23 & 716.0 & 0.25 & 0.042725893 & 0.060243509 & 0.080463381  \\ 
        WD 1101+ 364 & 0.36 & 0.31 & 97.0 & 0.16 & 0.404291575 & 0.570051121 & 0.7644969  \\ 
        WD 0931+4445 & 0.32 & 0.14 & 660.0 & 1.67 & 13.93049078 & 19.64199201 & 22.53379912  \\ 
        WD 1242-105 & 0.56 & 0.39 & 39.0 & 0.19 & 2.76990409 & 3.905564767 & 5.231591232  \\ 
        J0056-0611 & 0.174 & 0.46 & 585.0 & 0.53 & 1.194042867 & 1.683600443 & 2.201126945  \\ 
        J0106- 1000 & 0.191 & 0.39 & 2691.0 & 0.85 & 0.876060914 & 1.235245889 & 1.561517523  \\ 
                J0345+1748d & 0.76 & 0.181 & 166.0 & 0.1 & 0.074249388 & 0.104691638 & 0.140651886  \\ 
        J0745+ 1949d & 0.1 & 0.156 & 270.0 & 0.21 & 0.05777962 & 0.081469264 & 0.109032903  \\ 
        J0751-0141 & 0.97 & 0.194 & 1859.0 & 0.29 & 0.144504936 & 0.20375196 & 0.271508255  \\ 
        J0825+ 1152d & 0.49 & 0.287 & 1769.0 & 0.4 & 0.306112419 & 0.431618511 & 0.570704449  \\ 
        J1053+5200 & 0.26 & 0.213 & 1204.0 & 0.54 & 0.465211102 & 0.655947654 & 0.856771277  \\ 
        J1054-2121 & 0.39 & 0.168 & 751.0 & 0.22 & 0.076177336 & 0.107410044 & 0.143682038  \\ 
        .J1056+ 6536 & 0.34 & 0.338 & 1421.0 & 0.53 & 0.690171093 & 0.973141241 & 1.272277765  \\ 
        .J1108+ 1512 & 0.42 & 0.167 & 698.0 & 0.19 & 0.058355556 & 0.082281334 & 0.110217685  \\ 
        J1112+1117 & 0.14 & 0.169 & 257.0 & 0.13 & 0.024068587 & 0.033936708 & 0.045557547  \\ 
        J1130+ 3855 & 0.72 & 0.286 & 662.0 & 0.15 & 0.080363363 & 0.113312341 & 0.152016741  \\ 
        J1436+ 5010 & 0.46 & 0.233 & 830.0 & 0.51 & 0.987366444 & 1.392186686 & 1.823524315  \\ 
        J1443+ 1509 & 0.84 & 0.181 & 540.0 & 0.12 & 0.039921919 & 0.056289906 & 0.075586649  \\ 
        J1630+ 4233 & 0.3 & 0.307 & 820.0 & 0.84 & 3.394413377 & 4.786122861 & 6.057125911  \\ 
        J1741 + 6526 & 1.11 & 0.17 & 936.0 & 0.38 & 0.573272106 & 0.80831367 & 1.070443401  \\ 
        J1840+ 6423 & 0.65 & 0.177 & 676.0 & 0.12 & 0.025887731 & 0.0365017 & 0.049014848  \\ 
        J2338-2052 & 0.15 & 0.263 & 1295.0 & 0.3 & 0.067240209 & 0.094808695 & 0.126257818  \\ 
        CSS 41177 & 0.36 & 0.31 & 473.0 & 0.24 & 0.244447101 & 0.344670412 & 0.460599618  \\ 
        J1152 +0248 & 0.47 & 0.41 & 464.0 & 0.23 & 0.350741252 & 0.494545166 & 0.661224238  \\ 
    \end{tabular}
\end{threeparttable}
\end{table}

\nopagebreak

\end{document}